\documentclass[twocolumn]{aastex62}
\graphicspath{{./}{figures/}}
\usepackage{latexsym} 
\usepackage{graphicx}
\usepackage{amssymb}
\usepackage{longtable}
\usepackage{epsf}
\usepackage{lipsum} 
\usepackage{amsmath}
\usepackage[normalem]{ulem} 

\usepackage{xcolor} 
\usepackage{wasysym}
\usepackage{array,booktabs,ragged2e}
\newcolumntype{R}[1]{>{\RaggedLeft\arraybackslash}p{#1}}
\received{May 23, 2018}
\revised{January 9,2019}
\accepted{February 19, 2019}
\submitjournal{ApJ}

\shorttitle{Revised manuscript}
\shortauthors{Nikolaou et al.}

\begin{document}

\title{What factors affect the duration and outgassing of the terrestrial magma ocean?}

\correspondingauthor{Athanasia Nikolaou}
\email{athanasia.nikolaou@dlr.de}

\author[0000-0002-0786-7307]{Athanasia Nikolaou}
\affil{Institute of Planetary Research, German Aerospace Centre (DLR) \\
Rutherfordstr. 2, 12489 Berlin}
\affiliation{Centre of Astronomy and Astrophysics, Berlin Institute of Technology \\
Hardenbergstr. 36, 10623 Berlin}




\author{Nisha Katyal}
\affil{Institute of Planetary Research, German Aerospace Centre (DLR) \\
Rutherfordstr. 2, 12489 Berlin}
\affiliation{Centre of Astronomy and Astrophysics, Berlin Institute of Technology \\
Hardenbergstr. 36, 10623 Berlin}


\author{Nicola Tosi}
\affil{Institute of Planetary Research, German Aerospace Centre (DLR) \\
Rutherfordstr. 2, 12489 Berlin}
\affiliation{Centre of Astronomy and Astrophysics, Berlin Institute of Technology \\
Hardenbergstr. 36, 10623 Berlin}

\author{Mareike Godolt}
\affil{Centre of Astronomy and Astrophysics, Berlin Institute of Technology \\
Hardenbergstr. 36, 10623 Berlin}

\author{John Lee Grenfell}
\affil{Institute of Planetary Research, German Aerospace Centre (DLR) \\
Rutherfordstr. 2, 12489 Berlin}

\author{Heike Rauer\affil{1,2,3}}
\affil{Institute of Planetary Research, German Aerospace Centre (DLR) \\
Rutherfordstr. 2, 12489 Berlin}
\affiliation{Centre of Astronomy and Astrophysics, Berlin Institute of Technology \\
Hardenbergstr. 36, 10623 Berlin}
\affiliation{Department of Planetary Sciences, Institute of Geosciences, Free University of Berlin \\
Malteserstr. 74-100, 12249 Berlin}



\begin{abstract}

The magma ocean (MO) is a crucial stage in the build-up of terrestrial planets. Its solidification and the accompanying outgassing of volatiles set the conditions for important processes occurring later or even simultaneously, such as solid-state mantle convection and atmospheric escape. To constrain the duration of a global-scale Earth MO we have built and applied a 1D interior model coupled alternatively with a grey H$_2$O/CO$_2$ atmosphere or with a pure H$_2$O atmosphere treated with a line-by-line model described in a companion paper by \citet{Katyal2017}.
We study in detail the effects of several factors affecting the MO lifetime, such as the initial abundance of H$_2$O and CO$_2$, the convection regime, the viscosity, the mantle melting temperature, and the longwave radiation absorption from the atmosphere.
In this specifically multi-variable system we assess the impact of each factor with respect to a reference setting commonly assumed in the literature. We find that the MO stage can last from a few thousand to several million years. 
By coupling the interior model with the line-by-line atmosphere model, we identify the conditions that determine whether the planet experiences a transient magma ocean or it ceases to cool and maintains a continuous magma ocean. We find a dependence of this distinction simultaneously on the mass of the outgassed H$_2$O atmosphere and on the MO surface melting temperature. We discuss their combined impact on the MO's lifetime in addition to the known dependence on albedo, orbital distance and stellar luminosity and we note observational degeneracies that arise thereby for target exoplanets.

\end{abstract}


\section{Introduction}\label{intro}

Immediately after terrestrial planets form, their internal thermal structure is not well-known. This information is nevertheless important in order to understand the evolution of the atmosphere and the onset and development of mantle convection. 
 

With our numerical model we simulate the magma ocean (MO) phase, an intermediate stage in thermal evolution between completed accretion and the formation of a young planet's surface. During this period whose duration we aim to estimate, a large part of the mantle was fully molten.

Although the existence of the MO on the Moon is a widely accepted hypothesis due to its primary anorthositic crust \citep{Canup2004a,Sleep2014,Barboni2017}, such observational evidence remains elusive for the Earth. However, the Moon forming impact is expected to have molten extensively the Earth's mantle \citep[e.g.][]{Nakajima2015}. In addition, the release of gravitational potential energy associated with core formation, along with the kinetic energy of accretional impacts and the decay of short-lived radiogenic elements provide enough energy to globally melt the silicate mantle of an Earth-sized planet leading to the formation of a magma ocean \citep{Coradini1983,Solomatov2007,Elkins-Tanton2008,Sleep2014}. 

During the MO stage, the interior temperature is high and degassing of volatiles accompanies the thermal evolution \citep{AbeMatsui88,Elkins-Tanton2008,Zahnle2010,Schaefer2010,Lebrun2013,Gaillard2014,Massol2016,Salvador2017}. As a result a secondary atmosphere forms and is expected to provide the bulk of the atmospheric mass during the Hadean. Nevertheless, there is uncertainty in the initial volatile inventory of the Earth because impactors stochastically deliver volatiles to the accreting planets \citep{Morbidelli2005,Tsiganis2005,Raymond2017}.
The most conservative estimate of the resulting Earth's volatile budget considers 300 bar H$_2$O, roughly corresponding to today's ocean, and 100 bar CO$_2$, i.e. the amount estimated to be stored in the crust in the form of carbonates \citep{Ingersoll2013}. However, the existence of abundances in H$_2$O and CO$_2$ higher than those directly observable in the present Earth cannot be excluded and is even likely \citep{Hirschmann2009,Hirschmann2012,Hallis2015}. The bulk of the atmosphere outgassed from the magma ocean is suggested to be composed largely of H$_2$O and CO$_2$ for most carbonaceous chondritic materials \citep{Schaefer2010,Lupu2014} along with reduced species (H,CO,CH$_4$) \citep{Gaillard2014,Lupu2014}, excluding the case of chondritic binary mixtures \citep{Schaefer2017}. Proxy evidence for less oxidized early atmosphere is also given by sulphur isotope studies \citep{Ueno2009,Endo2016}.

As soon as a certain minimum amount of water vapor is present in the atmosphere, its role on the planetary evolution is the most crucial of all gases present. This is due to its strong greenhouse effect and the well-studied runaway greenhouse regime associated with it, which does not allow for radiative equilibrium solutions over a wide range of surface temperatures, while the atmospheric outgoing radiation stalls to a constant value known as the Kombayashi-Ingersoll (KI) or runaway greenhouse (RG) limit \citep[e.g.][]{Naka92,Kasting1988,Zahnle1988,Pierrehumbert2010,Leconte2013}. The RG regime over a magma ocean, however, would occur at surface temperatures in excess of 3000 K that are characteristic of a molten silicate mantle \citep{Lupu2014,Massol2016}. Previous research has mostly indicated an invariant outgoing radiation limit of $\sim$300 W/m$^2$ for an Earth-sized planet with water--dominated atmosphere \citep{Kasting1988,Zahnle1988,Naka92,Zahnle2007emergence,Kopparapu2013, Leconte2013,Hamano2015,Goldblatt2015_waterworlds}. 

In particular, \citet{Hamano2013}, extended the grey atmospheric model of \citet{Naka92} to cover high surface temperatures, and introduced the separation of the magma-ocean stage into short-term and long-term. Based on the comparison of the stellar irradiation to the KI limit of a H$_2$O-dominated atmosphere, they brought the role of stellar luminosity into context of the magma ocean lifetime, which, under suitable conditions, can hinder the planetary cooling altogether. However, in the \citet{Hamano2013} study, the potential role of mantle composition was not investigated \textbf{in combination with} the explicit role of the surface vapor pressure on the longwave radiation limit. We extend this previous work by considering such factors. 

The volatiles that envelop the terrestrial planets are crucial since they quantify the effect of thermal blanketing that delays radiative cooling of the MO by hundreds of thousands of years. It has been investigated by several authors so far \citep{AbeMatsui88,Elkins-Tanton2008,Lebrun2013,Hamano2013,Salvador2017,Hier2017,Ikoma2018}. However, comparing results from the literature is not straightforward because each magma ocean study involves many ad hoc assumptions. This is inevitable since different research fields focus on a specific niche of the MO system. At the same time, the topic is becoming increasingly multidisciplinary \citep{Tasker2017} and more of the assumptions are being challenged. 

Our aim is to calculate the lifetime of the magma ocean and to comprehensively assess the role of various parameters on it. Knowing their relative significance can help guide future model development. Therefore, we primarily account for the key role of the outgassed atmosphere. 
For planets which are volatile-poor or may quickly lose their atmosphere, blackbody thermal evolution is modeled and discussed. The findings are focused on an Earth-sized rocky planet. Yet their applicability in exoplanetary context is discussed and points of interest for the community are suggested.

 





\section{Methods}\label{section:methods}


\subsection{Numerical model}

We calculate the thermal state of the solidifying magma ocean without examining the preceding stage of its formation. We build a model that simulates the coupled evolution of the interior (Sections \ref{section:structure}-\ref{section:outgassing}.) and the atmosphere (Sections \ref{section:secondaryatm}-\ref{section:EarlySun}). The COnvective Magma Radiative Atmosphere and Degassing (COMRAD) model resolves the mantle interior profiles of temperature, liquid and solid fraction along with the degassing process, starting from a fully molten mantle up to the end of the magma ocean phase (see Section \ref{section:endofMO}).
We employ the mantle-surface temperature iteration method developed by \citet{Lebrun2013}, with differences in the calculation of the mantle adiabat (Section \ref{section:adiabat}), of the liquid viscosity (section \ref{section:viscosity}) and of the volatile mass balance (Section \ref{section:outgassing}). The outgassing of H$_2$O and CO$_2$ is calculated according to a melt solubility curve for each volatile (Section \ref{section:outgassing}). The atmosphere is treated in two alternative ways (Section \ref{section:secondaryatm}): i) A grey atmosphere accounting for two greenhouse gases H$_2$O and CO$_2$ \citep{AbeMatsui85,Elkins-Tanton2008} (Section \ref{section:greyatm}) and ii) a pure H$_2$O atmosphere with a spectrally-resolved Outgoing Longwave Radiation at the Top Of the Atmosphere (henceforth named ``OLR at TOA'' OLR$_\textrm{TOA}$) by \citet{Katyal2017} in a companion paper (hereafter ``companion paper'') (Section \ref{section:lblatm}). In the following sections we separately introduce each model component.
\ 


\subsection{Structure of the interior}\label{section:structure}



We consider a spherically symmetric Earth with outer radius $R_p$ and core radius $R_b$. This yields a mantle of thickness $R_p-R_b$ whose physical properties are defined by the melting curves (solidus and liquidus) of KLB-1 peridotite (Fig. \ref{fig:profiles}). 
By comparing the interior thermal profile to these curves, we identify the phase (liquid, partially molten, or solid) of different mantle layers. The mantle is initially assumed to be fully molten and convecting, with an adiabatic temperature profile. As it cools, the liquid adiabat (dotted line in Fig. \ref{fig:profiles}) intersects with the melting curves (solid lines in Fig. \ref{fig:profiles}). Due to the steeper slope of the adiabat compared to the melting curves, the adiabat and the liquidus intersect first atop the core-mantle boundary (CMB). The mantle is fully molten from the surface until the depth of intersection between the liquid adiabat and the liquidus; it is partially molten between the liquidus and solidus, and completely solid below the solidus. The melt fraction $\phi$ at any given depth, is calculated as \citep[e.g.][]{Solomatov1993a, Solomatov1993b, Abe97, Solomatov2007, Lebrun2013}: 
\begin{equation}
	\phi=\frac{T - T_{sol}}{T_{liq}-T_{sol}} \label{eq:phi}
\end{equation}
where $T$ is the temperature of the mantle at a given depth, and $T_{sol}$ and $T_{liq}$ the corresponding solidus and liquidus temperature.
The partially molten region is further divided by comparing the melt fraction $\phi$ with the critical melt fraction $\phi_C$ of 40\% that separates the liquid-like from solid-like behavior \citep{Costa2009}. For $\phi > \phi_C$, the region is considered liquid-like and belongs to the magma ocean convecting domain of depth $D$. The interface of phase change that separates the two regimes is called solidification or rheology front.
Note that $\phi_C$ varies among 30\% \citep[e.g.][]{Maurice2017,Hier2017}, 40\% \citep{Solomatov2007,Bower2017} and 50\% \citep{Monteux2016,Ballmer2017} in the geodynamic literature.



\subsection{Melting curves}\label{section:meltingCurves}

We use solidus and liquidus curves of KLB-1 peridotite obtained from experimental data. Depending on pressure, we adopt different parameterizations for different parts of the mantle. For the solidus, we use data from \citet{Hirschmann2000} for $P \in [0, 2.7)$ GPa, \citet{Herzberg2000} for $P \in [2.7, 22.5)$ GPa, and \citet{Fiquet2010} for $P \geq 22.5$ GPa, while for the liquidus, from \citet{Zhang1994} for $P \in [0, 22.5)$ GPa, and \citet{Fiquet2010} for $P\geq 22.5$ GPa. Since we employ data from multiple studies, we refer to the resulting set of melting curves as ``synthetic''. Such curves are adopted in our experiments unless otherwise specified. 

\begin{figure*}[htb!]
\centering
\includegraphics[width=0.8\textwidth]{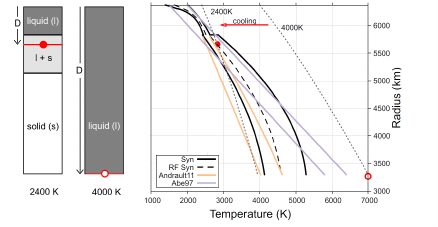}
\caption{Melting curves for three cases: linear according to \citet{Abe97} (``Abe97'', \textbf{purple solid lines}); synthetic for peridotitic composition according to \citet{Herzberg2000}, \citet{Hirschmann2000} and \citet{Zhang1994} for the upper mantle, and \citet{Fiquet2010} for the lower mantle (``Syn'', \textbf{black solid lines}); for chondritic composition according to the same data for the upper mantle and \citet{Andrault2011} for the lower mantle (``Andrault11'', \textbf{yellow solid lines}). ``Syn'' and ``Andrault11'' differ only in the lower mantle parametrization. The \textbf{black dashed line} indicates the profile of the rheology transition for the ``Syn'' curves (``RF Syn''). \textbf{Dotted lines} indicate adiabats with potential temperatures of 4000 K and 2400 K. The \textbf{red open} and \textbf{full circles} indicate the base of the liquid-like magma ocean of thickness $D$ for the two adiabats, with the corresponding depth ranges of liquid (l), solid (s), and partially molten (l+s) regions shown in the left columns.}
\label{fig:profiles}
\end{figure*}

As shown in Fig. \ref{fig:profiles}, we also tested the linear melting curves adopted by \citet{Abe97} and later by \cite{Lebrun2013}, as well as those introduced by \citet{Andrault2011} that are representative of a chondritic composition (for analytical expressions for all the melting curves see Appendix \ref{appendix:meltCurves}). Apart from the linear curves of \citet{Abe97}, the experimental data that we considered require higher order polynomial fittings since their slope is not constant with depth. As we discuss in Section \ref{section:adiabat}, this imposes a limitation in calculating the two-phase adiabat.

\subsection{Adiabat}\label{section:adiabat}

The interior temperature profile is calculated using the expression of the adiabatic temperature gradient for a one-phase system:
\begin{equation}
  \frac{dT}{dP} = \frac{\alpha_T T}{\rho c_P}, \label{eq:adiabat}
\end{equation}
where $P$ is the pressure in GPa, $c_P$ the thermal capacity at constant pressure, and $\alpha_T$ the pressure-dependent thermal expansivity given by \citep{Abe97}:
\begin{equation}
 \alpha_T(P) = \alpha_0\left[\frac{PK'}{K_0} \right]^{-(m-1+K')/K'}, \label{eq:aTofP}
\end{equation}
where $\alpha_0$ is the surface expansivity, $K_0$ the surface bulk modulus and $K'$ its pressure derivative (see Table \ref{tableConstants} for their values), and $m=0$. The pressure is simply calculated assuming a hydrostatic profile. 




\citet{Solomatov1993a,Solomatov1993b} derived the adiabat for a two-phase system by introducing a modified thermal expansivity and thermal capacity that depends on the melt fraction. However, the expressions they derived are explicitly valid for a system with constant rate of temperature drop with depth, which is equivalent to constant phase boundary slope and thus applies only to linear melting curves such as those of \citet{Abe97}. The modified adiabat then tends to align with the slope of constant melt fraction. Since they do not cover the higher order parameterizations of the experimental data that we adopted, we employ instead the one phase adiabat of Eq. \eqref{eq:aTofP} with constant thermal capacity and pressure-dependent expansivity. 

\subsection{Energy conservation and parametrized cooling flux}\label{section:energy}



Assuming that the mantle temperature profile $T(r)$ is adiabatic as described in Section \ref{section:adiabat}, the time-evolution of the magma ocean is obtained by integrating the energy-balance equation \citep{Abe97} over the evolving magma ocean volume:
\begin{equation}
\rho\left(c_P + \Delta H\frac{d\phi}{dT}\right) \frac{dT}{dt} = -\frac{1}{r^2}\frac{\partial(r^2 F_{conv}) }{\partial r} + \rho q_{r}, \label{eq:energy}
\end{equation}
where $\rho$ is the density, $\Delta H$ the specific enthalpy difference due to phase change, $F_{conv}$ the MO convective cooling flux, and $q_r$ the internal heat released by the decay of the radioactive elements (see Appendix \ref{section:Qrad}). 

Because of its large depth extent and liquid-like viscosity which approaches that of water (see Sect. \ref{section:viscosity}), the magma ocean is expected to undergo highly turbulent convection \citep{Solomatov2007} that is neither attainable in the laboratory \citep{Shishkina2016}, nor is numerically resolvable \citep{Hansen2015}. Key to the evolution of the interior temperature profile is the parametrization of the convective heat flux $F_{conv}$ of Eq. \eqref{eq:energy}. This is calculated with the aid of the Rayleigh ($Ra$) and the Prandtl ($Pr$) numbers: 


\begin{equation}
  Ra=\frac{\rho \alpha_T g (T_{p}-T_{surf})D^3}{\kappa_T \eta}, \quad Pr=\frac{\eta}{\rho\kappa_T} \label{eq:Ra} 
\end{equation}
%
%
where 
$g$ the gravity acceleration, $T_p$ the mantle potential temperature, $T_{surf}$ the surface temperature, $D$ the depth of the convective layer, $\kappa_T$ the thermal diffusivity, and $\eta$ the dynamic viscosity. 

We consider two different parameterizations for ``soft'' and ``hard'' turbulence. In the first one turbulent dissipation at the boundary layers affects the heat flux, while the flow is laminar in the bulk of the fluid \citep{Solomatov2007}: 
\begin{equation}
  F_\text{soft}=0.089 \frac{k_T (T_p-T_{surf})Ra^{1/3}}{D} \label{eq:Fsoft}
\end{equation}
where $k_T = \kappa_T\rho c_p$ is the thermal conductivity. With the above formulation, the heat flux becomes independent of the depth of the convective layer since the Rayleigh number is proportional to $D^3$. The second parameterisation depends additionally on the inverse of the Prandtl number. It represents a regime where the heat flux is assumed to be controlled not only by boundary layer friction but also has a contribution from turbulence generated in the bulk volume of the fluid \citep{Solomatov2007}: 

%
\begin{equation}
  F_\text{hard}=0.22\frac{k_T (T_{p}-T_{surf})}{D}Ra^{2/7} Pr^{-1/7} \lambda^{-3/7}\label{eq:Fhard}
\end{equation}
where $\lambda$=basin Length/Depth is the aspect ratio of the mean flow. 
For very high $Ra$, the hard turbulence parameterization is suggested \citep{Solomatov2007}. There, increasing values of $Pr$ in a progressively more viscous fluid yields lower heat flux if all other parameters are left unchanged. Both parameterizations were implemented and tested.

\subsection{Magma ocean viscosity}\label{section:viscosity}


The melt viscosity prominently factors into the thermal evolution of the magma ocean (Eq. \ref{eq:Ra}, \ref{eq:Fsoft}). We separate the mantle into two regimes, namely liquid-like and solid-like. The transition between them upon cooling and solidification is a complex phenomenon that depends, among other factors, on composition and cooling rate \citep{Speedy2003}. 
While the viscosity dependence on temperature follows an Arrhenius law below the solidus temperature \citep{Kobayashi2000}, non-linear effects take place near and above it \citep{Dingwell1996, Kobayashi2000, Speedy2003}. 
In order to mitigate the solid state transition \citet{Salvador2017} proposed a ``smoothening'' of the sharp viscosity jump that occurs in earlier magma ocean models \citep[e.g.][]{Lebrun2013,Schaefer2016}. However, during cooling the crystal content increases and the melt rheology is expected to make a discontinuous jump from the liquid- to solid-like state over a short crystallinity range \citep{Marsh81}. Continuous variation in viscosity across 5 orders of magnitude is seen only in the case of glasses \citep{Kobayashi2000}, although such behaviour is not consistent with common MO solidification assumptions.  

The Vogel-Fulcher-Tammann, henceforth referred to as VFT equation is employed in our work. By VFT definition, the obtained viscosity tends to an infinite (hence not physically meaningful) value at a threshold temperature $T=C$. 
%
%
We consider two such expressions for the liquid dynamic viscosity: one that depends only on the temperature $\eta_l=f(T)$ \citep{Karki2010} and a second one that depends on both temperature and water content $\eta_l=f(T,X_{H_2O})$ \citep{Giordano2008}. 

\citet{Karki2010} found that for hydrous melts, the viscosity at a given potential temperature can be well fitted to the following VFT equation:
\begin{equation}
  \eta_l(T)=A_K \exp\left[\frac{B_K}{T - C_K}\right], \label{eq:nliqKarki}
\end{equation} 
where $A_K$, $B_K$, $C_K$ (See Table \ref{tableConstants} for their values) are calculated for a fixed water content of 10 wt\%. 

\begin{figure}[ht!]
\hspace{-0.5cm}
\includegraphics[scale=0.4]{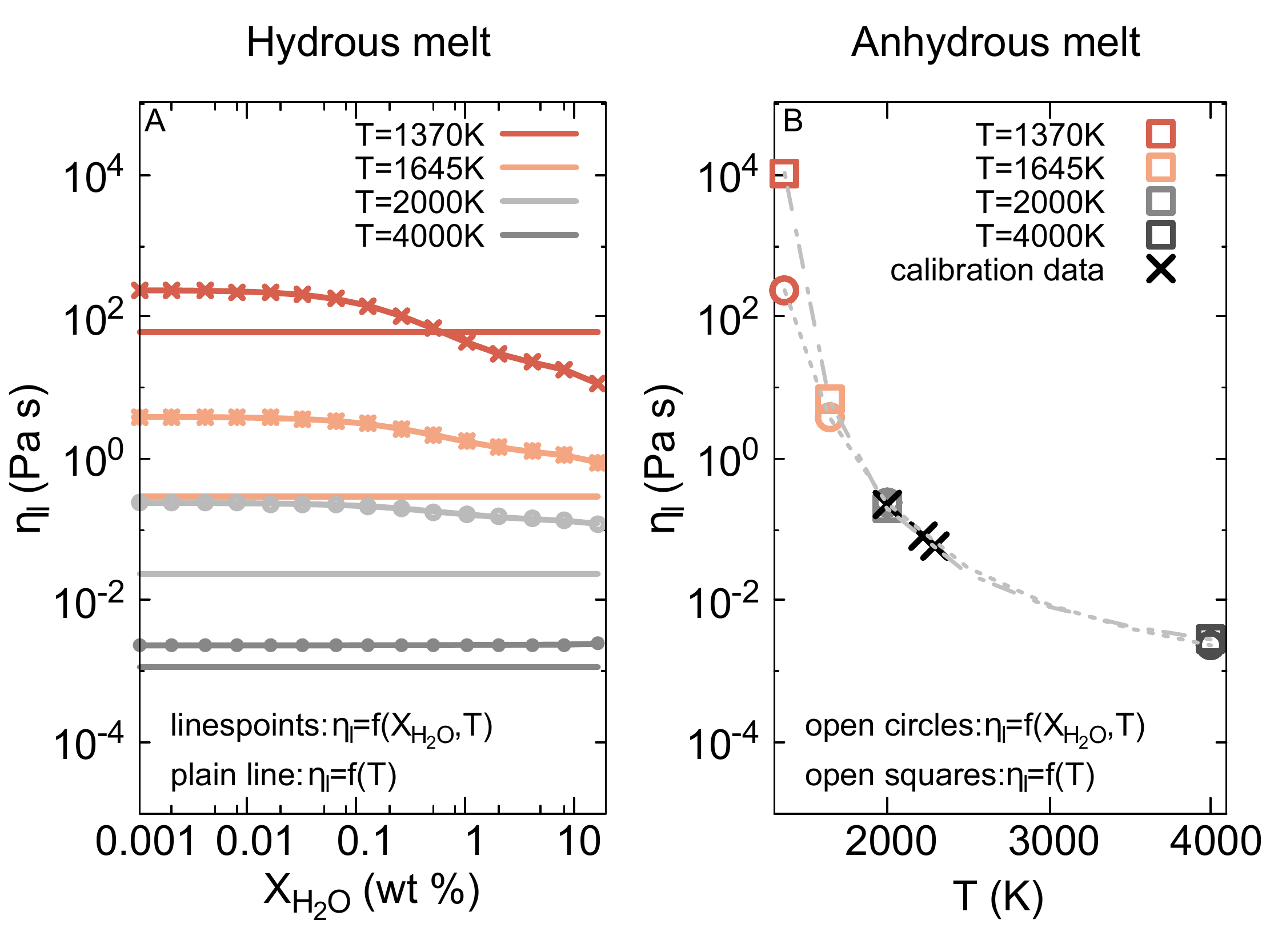}
\caption{Variation of melt dynamic viscosity $\eta_l$ with temperature for hydrous and anhydrous melt. \textbf{A:} Melt viscosity as a function of water content for different temperatures. Equation \eqref{eq:nliqKarki}, which assumes a fixed water concentration of 10 wt\% \textbf{(solid lines)}, while Eq. \eqref{eq:nliqGiordano} explicitly includes the effect of water concentration \textbf{(linepoints)}. \textbf{B:} Viscosity of anhydrous melt as a function of temperature. \textbf{Squares} and \textbf{circles} are obtained with Eq. \eqref{eq:nliqKarki} and Eq. \eqref{eq:nliqGiordano} with $X_{H_2O}=0$, respectively. 
Experimental values of anhydrous silicate melt are obtained from \citet{Urbain82} (see Appendix \ref{appendix:viscCalibration}).}
\label{fig:viscosityVFT}
\end{figure}

However, Eq. \eqref{eq:nliqKarki} does not explicitly include the effect of water content, which is expected to vary during the simulation (see Section \ref{section:evolutionoutgassing}). The presence of water tends to lower the melt viscosity \citep{Marsh81,Dingwell1996,Giordano2008,Karki2010} and it is important to include it as a time-dependent variable in our calculations. 
 Hence, we implemented the empirical model of \citet{Giordano2008} 
which uses explicitly the water concentration, together with the concentration of 13 different oxides in the silicate melt. 
It calculates two of the three VFT parameters ($B_G$ and $C_G$ in Eq. \eqref{eq:nliqGiordano} below). The viscosity for a given temperature $T$ and water concentration $X_{H_2O}$ is then given by:
\begin{equation}
  \eta_l(T,X_{H_2O})=10^{A_G +\left[\frac{B_G}{T - C_G}\right]}, \label{eq:nliqGiordano}
\end{equation} 
where the parameter $A_G=-4.55 \pm 1$ is a constant pre-exponential factor. The parameters $B_G(X_{H_2O})$ and $C_G(X_{H_2O})$ are calculated by the model at each time step according to the evolving concentration of water (see Section \ref{section:outgassing}). Even though COMRAD is unable to resolve the evolution of the melt composition, it does resolve the melt water concentration with time, which allows us to evaluate the \citet{Giordano2008} model at each time step. In order to use this model, it is required that we choose a suitable composition as a constant, non-evolving basis. We found that the composition of basanite \citep{Giordano2003, Giordano2008} is able to reproduce the experimental values for the temperature dependent viscosity for the anhydrous case \citep{Urbain82} since it is one of the least evolved in terms of silicate content. After calibration of the prefactor $A_G$ (see Appendix \ref{appendix:viscCalibration}, Table \ref{table:nliq_calibration}), the error is estimated to lie within $\pm$ 10$\%$. This fitting provides us with a parameterized description of the composition that allows us to treat the decrease of the melt viscosity with increasing water content (Fig. \ref{fig:viscosityVFT}A). 
For anhydrous melt, the viscosity calculated with Eq. \eqref{eq:nliqGiordano} and $X_{H_2O}=0$ yields similar values as those proposed by \citet{Karki2010} at high temperatures, though the two tend to depart significantly for temperatures near the solidus (Fig. \ref{fig:viscosityVFT}B). 

The  pressure dependence of the viscosity for the hydrous case is not explicitly provided by \citet{Karki2010}. 
The authors report that it varies by a relatively small factor between 2.5 and 10 over the pressure range [0, 140] GPa spanned by a global terrestrial magma ocean. 
In the following, for simplicity, we will neglect such dependence and use the liquid viscosity evaluated at the potential temperature $T_p$ of the magma ocean as representative for the fully molten part.

The melt viscosity $\eta_{l}$ is further corrected for the crystal fraction content in each layer. In the liquid-like partially molten region, the effect of crystals is taken into account with the following expression \citep{Roscoe52}:
\begin{equation}
  \eta = \frac{\eta_l}{\left(1- \frac{1-\phi}{1-\phi_C}\right)^{2.5}} \label{eq:nliqphi}
\end{equation}
By combining Eq. \eqref{eq:nliqKarki} or \eqref{eq:nliqGiordano} with Eq. \eqref{eq:nliqphi} for each layer that belongs to the magma ocean, we obtain the volumetric harmonic mean viscosity $\eta$ that is then used in the calculation of the parametrized convective heat flux in Eq. \eqref{eq:Fsoft}. The effect of the layer with the lowest viscosity value is prioritized in the calculation of harmonic mean viscosity and sets the leading order of magnitude for the value used in the calculation of the Rayleigh number \eqref{eq:Ra}.

The viscosity employed during the MO lifetime is the liquid-like viscosity ($\eta_l$). Note that the solid-like viscosity ($\eta_s$) is employed only after the MO phase ends and it is expressed after the \citet{Karato1993} formulation. The viscosity of the partially molten solid-like region below the rheology front (i.e. for $\phi<\phi_C$), is modified by the presence of the crystals according to \citet{Solomatov2007}. 

\subsection{Outgassing}\label{section:outgassing}

Along with the thermal evolution the concentration of volatiles in the mantle and their outgassing into the atmosphere is calculated. 
We use solubility curves to calculate the concentration and gas pressure in the melt for each volatile. For H$_2$O we use \citep{Caroll1994}:
\begin{equation}
    P_{sat,H_2O} = \left(\frac{X_{H_2O}}{6.8 \cdot 10^{-8}}\right)^{(1/0.7)},
    \label{eq:PsatH2O}
\end{equation}
and for CO$_2$ \citep{Pan1991}:
\begin{equation}
    P_{sat,CO_2} = \frac{X_{CO_2}}{4.4 \cdot 10^{-12}}.
    \label{eq:PsatCO2}
\end{equation}
In this way, we obtain the saturation vapor pressure over melt with a given volatile concentration. Due to the efficient mixing the volatile concentration is homogeneous throughout the magma ocean.

Upon solidification, part of the volatile budget remains into the solid mantle according to the partition coefficients of lherzolite for the upper mantle ($\kappa_{vol, lhz}$) and of perovskite for the lower mantle ($\kappa_{vol, pv}$) (see Table \ref{tableConstants} for their values). 
By calculating the volatile content stored in the liquid and solid phases of the mantle, we estimate the mass balance for each volatile at each time iteration $t$ as follows:
\begin{equation}
\begin{aligned}
M_{l,t_0}X_{vol,t_0}= & P(X_{vol,t}) \frac{4 \pi R_p^2}{g}  \\
& + M_{s,pv} \kappa_{vol,pv}X_{vol,t} + M_{s,lhz} \kappa_{vol,lhz}X_{vol,t} \\
    & + M_{l,z<z_{RF}}X_{vol,t} + M_{l,z>z_{RF}}X_{vol,t}\Pi  \\
    & + M_{l,z>z_{RF}}X_{vol,t-dt}(1-\Pi),
\end{aligned}
\label{eq:massBalance}
\end{equation}
%
where $M_{l,t_0}$ is the initial ($time=t_0$) mass of the liquid mantle, $X_{vol,t_0}$ the initial volatile concentration in the melt, $P(X_{vol,t}$) the saturation pressure of the volatile for the respective concentration $X_{vol,t}$ at time $t$, $M_{s,pv}$, $M_{s,lhz}$ is the mass of solid mantle in perovskite and in lherzolite respectively, $M_l$ the mass of the melt at depth $z$ either shallower ($z<z_{RF}$) or deeper than the rheology front ($z>z_{RF}$), and $X_{vol,t-dt}$ the concentration of the volatile in the previous time step. Comparing the melt percolation velocity \citep{Solomatov2007} to the rheology front velocity we calculate a volumetric fraction $\Pi$ of the total melt volume that upwells across the rheology front. $\Pi$ takes values within 0 and 1. 
The last term on the RHS of Eq. (\ref{eq:massBalance}) represents the volatile mass trapped in the liquid below the rheology front and is evaluated at the concentration of the previous time step.

Equation (\ref{eq:massBalance}) combined with either Eq. \eqref{eq:PsatH2O} or \eqref{eq:PsatCO2}, forms a system of two equations in two unknowns $(P,X)$ for each species. Eq. \eqref{eq:PsatH2O} is non-linear with respect to $X_{H_2O}$ and is solved iteratively with 0.01 bar tolerance. 

The thermal evolution of the system is coupled to the outgassing process. Upon cooling, the mantle volume is redistributed into a solid and liquid phase (see Fig. \ref{fig:profiles}). Consequently, the masses of the solid and liquid reservoirs $M_s$ and $M_l$ where the volatiles are stored change continuously. For every new mantle layer that solidifies volatile enrichment is ensured in the remaining melt. Since the saturation pressure increases monotonically with concentration (Eqs. \ref{eq:PsatH2O} and \ref{eq:PsatCO2}), the equilibrium gas pressure also increases, resulting in a progressive build-up of atmospheric mass at the surface of the planet. 

\subsection{Secondary atmosphere}\label{section:secondaryatm}

We adopt two alternative approaches to model the atmosphere generated upon magma ocean outgassing: i) a grey approximation after \citet{AbeMatsui85} that treats two gas species H$_2$O and CO$_2$ and ii) a line-by-line approach that calculates a spectrally resolved OLR (companion paper) that assumes a pure H$_2$O vapor composition. 

\subsubsection{Grey atmospheric model}\label{section:greyatm}

The grey approximation that we use is derived in \citet{AbeMatsui85}. It considers the absorption of thermal radiation independently of the wavelength. Both outgassed species H$_2$O and CO$_2$ absorb significantly in the spectral region where thermal energy is emitted from the surface of the Earth, and are therefore greenhouse contributors. By absorbing radiative energy, they exert a direct control on the surface temperature. Water is the most potent greenhouse agent of the two under normal atmospheric conditions ($P_0 = 101325$ Pa, $T_0 = 293$ K) \citep[see e.g.][]{Pierrehumbert2010}. For the H$_2$O absorption coefficient the value $k_{0,H_2O} = 0.01$ m$^2$/kg in the mid-infrared window region 1000 cm$^{-1}$ is adopted, after \citet{AbeMatsui88}. CO$_2$ is accounted for with absorption coefficient $k_{0,CO_2} = 0.001$ m$^2$/kg \citep{Yamamoto1952}. 
A higher value $k_{0,CO_2} = 0.05$ m$^2$/kg has been employed by \citet{Elkins-Tanton2008} (along with lower $k_{0,CO_2}$ values) and by \citet{Lebrun2013}, which was calculated by \citet{Pujol2003} in order to reproduce present day's Earth climate sensitivity (ECS). ECS refers to the combined response of the climate system to the radiative forcing from doubling the atmospheric CO$_2$ abundance relative to its pre-industrial levels and corresponds to an increase of about 2$^{\circ}$C in surface mean temperature \citep{IPCCFifth}. Using it is a good practice for studying the role of CO$_2$ radiative forcing on today's temperate Earth climate, within which the water vapor is not saturated over the atmospheric column. The presence of a liquid ocean is a strong constraint on the ECS and affects the overlying atmospheric profile through the inter-component exchange of vapor or ``hydrological cycle'' \citep{HeldSoden2006}, provided that no runaway greenhouse regime \citep{Pierrehumbert2010} ensues. Extrapolating the climate sensitivity to MO mean surface temperature ($>$1000 K) well above today's (300 K) is unsuitable for our study. 
We therefore avoid using the ECS-based value for $k_{0,CO_2}$ because it could overestimate CO$_2$'s radiative forcing on a planet with qualitatively different surface and atmospheric dynamics.
In \citet{AbeMatsui85} the downward radiation at the TOA is set to the incoming stellar flux $F_{Sun}$, which depends on the incident radiation $S_0$ at the assumed orbital distance. It relates to the blackbody equilibrium temperature $T_{eq}$ of the planet through the Stefan-Boltzmann law:
\begin{equation}
 F_{Sun}=\left(1-\alpha \right) \frac{S_0}{4}=\sigma T_{eq}^4,
  \label{eq:Teq}
\end{equation}
where $\alpha$ is the albedo and $\sigma$ the Stefan-Boltzmann constant. The resulting net upward flux at the top of the atmosphere ($F_{grey}$) is given by \citep{AbeMatsui85}:
%
\begin{equation}
 F_{grey}=\sigma \epsilon \left(T_{surf}^4 - T_{eq}^4\right)=F_{conv}
  \label{eq:Fgrey}
\end{equation}
%

According to Eq. \eqref{eq:Fgrey}, the net radiative flux at the TOA is positive for $T_{surf} > T_{eq}$. We adopt the convention of positive flux to represent planetary cooling. In order to find a state of the system that satisfies the energy balance, assuming that the radiative atmospheric adjustment is instantaneous, we require that the convective heat flux $F_{conv}$ at the top of the magma ocean is equal to the flux at the TOA. 

%
For a given potential temperature, we solve the system of Eq. \eqref{eq:Fsoft} and \eqref{eq:Fgrey} using an iterative scheme built according to the method of \citet{Lebrun2013} with an accuracy of $10^{-2}$ W/m$^2$. 


\subsubsection{Line-by-line atmospheric model}\label{section:lblatm}

An alternative approach is to employ a line-by-line (lbl) code to calculate the atmospheric outgoing radiation. The respective model is described in the companion paper. The advantage of this approach is that it provides a detailed calculation of wavelength-dependent longwave emission. We assume a 100$\%$ water vapor atmosphere (``steam atmosphere'') as commonly used by various authors when treating magma ocean planets or exoplanets with water--dominated atmospheres \citep[e.g.][]{Hamano2013, Massol2016, Schaefer2016}. A dry adiabatic temperature profile is used for the troposphere when the surface temperature is above the critical point of water ($T_{H2O,crit} = 647$ K). When the dry adiabat intersects the saturation vapor pressure of water, a moist adiabatic temperature profile is assumed \citep{Kasting1988}.


Using the temperature profiles of the atmosphere for a range of surface temperatures $T_{surf}$ and surface pressures $P_{H_2O}$, the emitted radiation is calculated for the spectral range from $20-29,995$ cm$^{-1}$ using the line by line model GARLIC \citep{Schreier2014} with HITRAN 2012 \citep{Rothman2013}. Integrating the emitted radiation at the TOA (corresponding to pressure 1 Pa), the outgoing radiation flux OLR$_{TOA}$ is obtained.

\citet{Katyal2017} determines OLR$_{TOA}$ for various H$_2$O surface pressures and temperatures on a ($P_{H_2O,0},T_{surf}$) grid that covers pressures between 4 and 300 bar and temperatures between 650 and 4000 K (Appendix \ref{appendix:LBL}; Fig. \ref{fig:Product}). For the surface vapor pressure we use the values 4, 25, 50, 100, 200, and 300 bar. For the surface temperature, we sample our calculations with a resolution $\Delta T= 100$ K. We employ $\Delta T= 20$ K in the region $T\in[1400, 2200]$ K where the highest variability of outgassing with respect to surface temperature occurs for the melting curves used in this study. 
We obtain values of OLR$_{TOA}$ that correspond to conditions intermediate to the grid points via bilinear interpolation. The relative interpolation error ranges from 10\% for fluxes of the order of 10$^6$ W/m$^2$ to about 1\% for fluxes of the order of 10$^2$ W/m$^2$.

In order to couple the lbl model results with the magma ocean thermal evolution, we impose a balance between the net energy flux at TOA and the magma ocean cooling flux $F_{conv}$ such that:
\begin{equation}
F_{conv}=OLR_{TOA}-F_{Sun}.
 \label{eq:FLBLBalance}
\end{equation}
%

Equations \eqref{eq:Fsoft} and \eqref{eq:FLBLBalance} form a system of two unknowns $T_{surf}$ and $F_{conv}$, which we solve iteratively with a tolerance of $10^{-1}$ W/m$^2$. 
The resulting flux needed to balance the RHS of Eq. \eqref{eq:FLBLBalance} can be either positive or negative, corresponding to a cooling or warming case, respectively.




\subsection{Incoming stellar radiation}\label{section:EarlySun}

For the young Sun we used a lower irradiation value following the expression for the time dependence of the solar constant of \citet{Gough1981}; otherwise we used today's value (1361 W/m$^2$, see Table \ref{table:Ref}).

Calculating the planetary albedo is outside the scope of this study. $\alpha$ is instead used as an input parameter. The suggested albedo for a cloudless steam atmosphere lies within the range [0.15, 0.40] \citep{Kasting1988,Goldblatt2013,Leconte2013,Pluriel2019}. We employ the value 0.30 unless otherwise stated. 

\subsection{End of the magma ocean phase}\label{section:endofMO}

The end of the magma ocean phase is defined as the point in time when the rheology front reaches the surface. At that stage, the mantle adiabat has potential temperature $T_{RF,0}$ such that all mantle layers have a melt fraction lower than $\phi_{C}$, a condition termed ``mush stage'' by \citet{Lebrun2013}. Although some melt still remains enclosed in the solid matrix, the mantle subsequently behaves as a solid. Moreover, the adiabatic profile used for the solid mantle implies that solid state convection has fully developed. While this is a likely scenario for slowly solidifying magma oceans, establishing whether and to what extent the solid mantle convects during its early stages is beyond the scope of this study since this would require the use of fully dynamic simulations \citep[e.g.][]{Maurice2017,Ballmer2017}. We therefore consider the established convection considered in this work to be an end member case within the geodynamic assumptions.

However, we stress that the thermal evolution model is designed to cover only the time until the MO end is reached via bottom-up crystallization. 

\begin{table*}[hbt]
\caption{Parameter values and components of the Reference-A model. The table is for complementary use to the results presented in Table \ref{table:sensitivity}.}
\centering
\renewcommand{\arraystretch}{0.8} 
\begin{tabular}{l c c c}
\hline
\hline
  Parameter  & Description & Value/Type  & Unit/Info  \\
 \hline
   Atmosphere & Type of approximation & grey  & Eq. \eqref{eq:Fgrey} \citet{AbeMatsui85}  \\
   $k_{0,H_2O}$  &  Absorption coeff. at normal atmospheric conditions & 0.01  & m$^2$/kg \\ 
   $k_{0,CO_2}$  &  Absorption coeff. at normal atmospheric conditions & 0.001 & m$^2$/kg \\ 
   H$_2$O content & Total water reservoir & 300 & bar   \\
   X$_{H_2O,0}$ & Inital H$_2$O mantle abundance & 410 & ppm   \\   
   CO$_2$ content & Total CO$_2$ reservoir  & 100 & bar \\
   X$_{CO_2,0}$ & Initial CO$_2$ mantle abundance & 130 & ppm \\
   $S$ & Solar constant $(S_0)$ &  1361 & W/m$^2$ \\
   $\alpha$  & Planetary albedo & 0.30 & -  \\
   $T_{p,0}$ & Initial potential temperature & 4000 & K  \\
   $D$ & Initial MO depth  & 2890 & km \\     
   $\eta_l$  & Melt viscosity parameterization & $\eta_l=f(T)$ & (Eq. \ref{eq:nliqKarki})   \\
   $q_{r}$  & Radioactive heating & 0 & not included \\
   $t_{planet}$ & Planet accretion time & 100 & Myr (employed for $q_r$ only) \\ 
   $T_{sol}$, $T_{liq}$ & Melting curves & ``synthetic'' & \citet{Herzberg2000,Hirschmann2000} \\
     & & & and \citep{Fiquet2010} (Section \ref{section:meltingCurves}) \\
   $T_{RF,0}$ & Temperature of rheology front at $z=0$  & 1645 & K \\
   $F_{conv}$ & Convective heat flux parameterization & soft turbulence & Eq. \eqref{eq:Fsoft} \citet{Solomatov2007} \\
 \hline
 \end{tabular}\label{table:Ref}
 \end{table*}

\section{Experimental setting}\label{section:refA}

Since our model has numerous input parameters, we define a set of parameter values, hereafter called ``Reference-A'' (Ref-A) model setting (reported in Table \ref{table:Ref}), with respect to which we perform changes and comparisons. This model is intended to be as straightforward as possible, to facilitate model comparison. It does not include radioactive heat sources, the melt viscosity only depends on temperature according to Eq. \eqref{eq:nliqKarki}, the abundance of volatiles is set to today's Earth observed reservoirs and it uses the atmospheric grey model for two species H$_2$O and CO$_2$. Additional aspects such as the solar irradiation and type of melting curves used are also defined.
For completeness, we note here that the suffix ``-A'' is necessary in order to mark a clear distinction to the ``Reference-B'' special setting that is used in Section \ref{section:sensitivity}. ``Ref-B'' differs from ``Ref-A'' in that it does not include CO$_2$. 

The experiments are organized as follows: We firstly examine the thermal and dynamical evolution of the magma ocean in the absence of an atmosphere and under the influence of grey /lbl atmosphere (Section \ref{section:evol3cases}). We examine the simultaneous evolution of H$_2$O and CO$_2$ outgassing, and vary the initial volatile abundances in order to calculate their effect on the magma ocean solidification time (Section \ref{section:evolutionoutgassing}). We quantify the minimum remnant volatiles in the mantle at the end of the magma ocean and we study the influence of the choice of melting curves on the evolution of water outgassing. Concluding the overview of the coupled interior-atmosphere system, we then study the separate influence of each parameter (or parameterized process) upon the solidification time (Section \ref{section:sensitivity}). 
In Sections \ref{section:greyVSlbl}--\ref{section:multiAlbedo} we shift our focus to the influence of the steam lbl atmosphere and use the atmospheric calculations of the companion paper. We show the qualitative difference between the grey and the lbl water vapor atmospheres (Section \ref{section:greyVSlbl}). We discuss the mechanism which separates the transient from the continuous magma ocean regime (Section \ref{section:mechanism}), and we find the critical albedo that separates the two, for an atmospheric water inventory at a constant distance from the star (Section \ref{section:orangeAlbedo}).
Finally, in Section \ref{section:multiAlbedo}, we expand the critical albedo calculation with dependence on the outgassed water vapor and the temperature of the rheology front at the surface. In Section \ref{section:MOtypes} we discuss the distinction between ``evolutionary'' and ``permanent'' magma oceans. The relevance of the results in the context of exoplanets is discussed in Section \ref{section:exoplanets} and a summarizing plot of the solidification time according to the factors examined is provided along with the Discussion.


\section{Results}\label{section:GREYresults}

\subsection{Thermal and dynamical evolution}\label{section:evol3cases}

\begin{figure*}[htb!]
\centering			
\includegraphics[width=0.8\textwidth]
{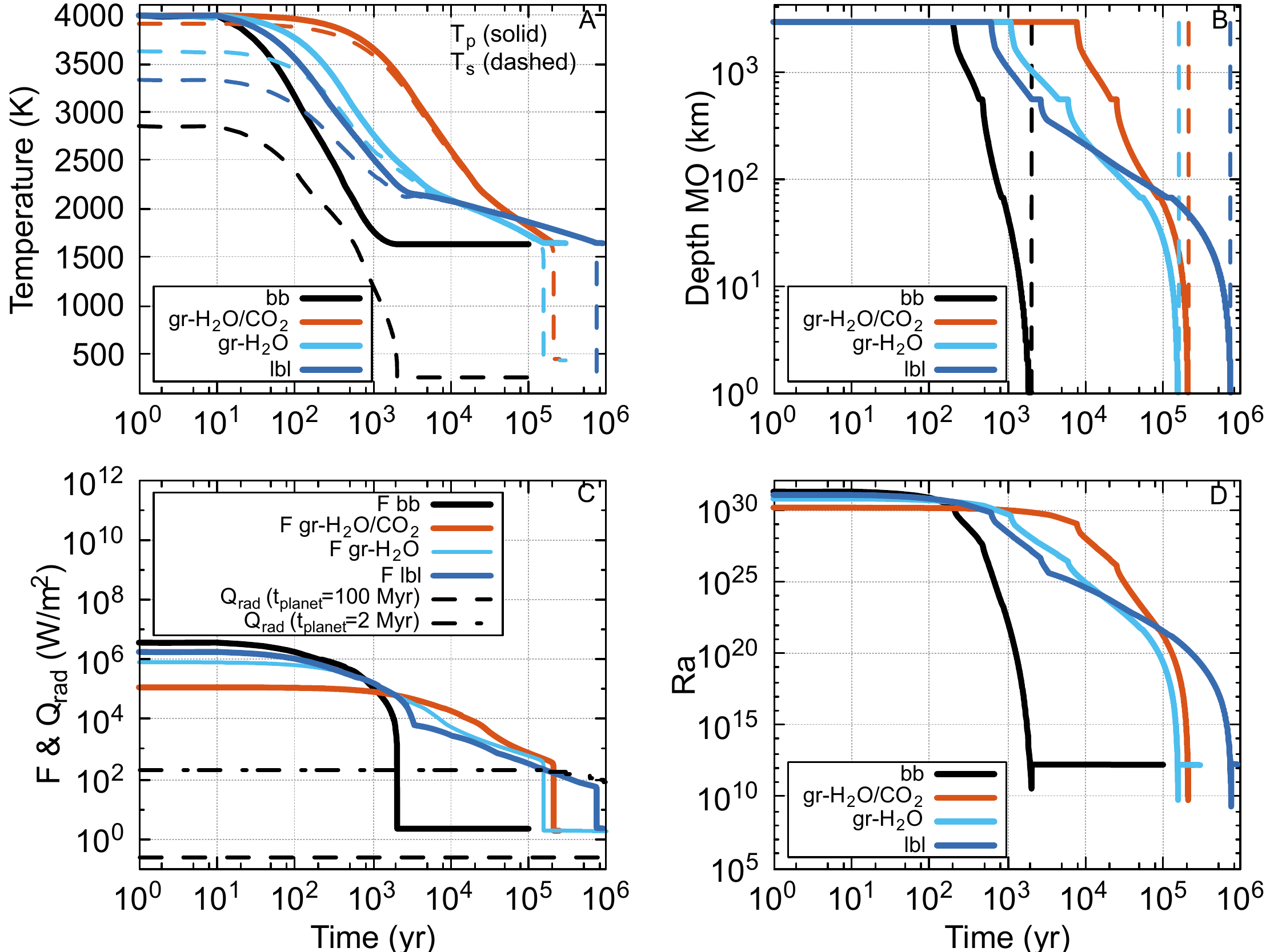}
\caption{Thermal evolution of black body (bb), grey H$_2$O atmosphere (gr-H$_2$O), grey H$_2$O/CO$_2$ atmosphere (gr-H$_2$O/CO$_2$), and line-by-line H$_2$O atmosphere (lbl). \textbf{A:} Evolution of potential (solid) and surface (dashed) temperature; \textbf{B:} Evolution of the depth of the magma ocean (dashed lines indicate the end of MO); \textbf{C:} Evolution of convective energy sink compared to the energy source of radioactivity. Note that the contribution of the radioactive heat sources is not included in the Ref-A settings and is only plotted for comparison; \textbf{D:} Evolution of $Ra$ number. Apart from the explicit differences in the atmospheric component, all other parameters are taken from the Ref-A case (Table \ref{table:Ref}).} 
\label{fig:evol3cases}
\end{figure*}

In a similar fashion to prior studies of the magma ocean solidification \citep{Zahnle1988,Abe97, Elkins-Tanton2008,Hamano2013,Hamano2015,Lebrun2013,Monteux2016,Schaefer2016,Hier2017} we present the thermal evolution using the state variables: surface temperature, potential temperature, heat flux, Ra number, and MO depth evolution. This enables both model validation and comparison. We adopt the multipanel approach of \citet{Lebrun2013} that is convenient for comparisons between varying modeling approaches. We performed four simulations for the following cases: i) absence of atmosphere referred to as the blackbody ``bb'' case, ii) a grey atmosphere composed of both H$_2$O and CO$_2$ ``gr-H$_2$O/CO$_2$'', iii) a grey atmosphere composed of only H$_2$O ``gr-H$_2$O'' and iv) a H$_2$O atmosphere treated with a line-by-line ``lbl'' model (Fig. \ref{fig:evol3cases}).
Apart from the representation of the atmosphere or absence thereof, all aspects of the model follow the Ref-A case (Table \ref{table:Ref}). 

Commonly in all simulations, the $T_p$ and $T_{surf}$ co-evolve until an abrupt difference between the two marks the end of the magma ocean (Fig. \ref{fig:evol3cases}B dashed lines); the liquid-like behavior comes to an end and a layer with a melt fraction of 40\% or lower remains. The average viscosity increases by more than 8 orders of magnitude across the critical melt fraction, taking values from 10$^8$ to 10$^{18}$ Pa$\cdot$s (not shown). A smooth variation across this interval would be difficult to justify under the assumption of fractional crystallization \citep[e.g.][]{Marsh81}. At this point, the whole domain switches to low cooling flux that characterizes solid-like convection (Eq. \ref{eq:Fsoft}) and the surface temperature drops abruptly, while the potential temperature remains unaffected.

The ``bb'' thermal evolution compares well with that presented by \citet{Lebrun2013}. It demonstrates the highest $T_p-T_s$ difference. The mantle consequently cools rapidly (0.002 Myr) with the highest convective flux ($F =5\cdot 10^6$--10$^4$ W/m$^2$), caused by this large temperature difference. Longer solidification times (0.15 Myr) are found by \citet{Monteux2016} who assume a bb radiative cooling ($F=10^5$--10$^2$ W/m$^2$) but a different interior model with a heat contribution from the core. The bb case is only relevant for planets that lose their outgassed atmosphere instantaneously.

For a planet that retains its atmosphere, the grey approximations show that the presence of the additional greenhouse species CO$_2$ contributes only 0.05 Myr to the solidification time and is less significant in comparison to the water (0.16 Myr vs 0.21 Myr MO duration). The longer solidification time ($\approx$0.4 Myr) obtained by \citet{Lebrun2013} for their grey two--species case is consistent with absorption coefficient $k_{0,CO_2}=0.05$, which is likely to be rather high for those climates (see Section \ref{section:greyatm}). The grey approach employed in this study and theirs follows \citet{AbeMatsui85} and should not be identified with other grey models used in the literature: The study of \citet{Hamano2013} expands on the \citet{Naka92} grey model and employs supercritical water thermal capacities. The study of \citet{Hier2017} formulates a hybrid energy balance for the atmosphere employing elements from both \citet{AbeMatsui85} and \citet{Hamano2013}. For potential temperature equal to the equilibrium temperature it results in net radiative warming of the planet. Moreover, the study of \citet{Hier2017} preserves the mantle fully molten for the majority of the MO period due to the lack of convective cooling sink. 
The slow solid-matrix compaction process provided from their detailed melt/volatile percolation model further increases the solidification time (3 Myr) in comparison to our study. Lastly, we obtain lower solidification time in comparison to \citet{Hamano2013} who define the MO end at the surface solidus and not at the higher temperature of the critical melt fraction.

The cooling path can be followed from the convective heat flux, the MO depth and the $Ra$ number (Fig. \ref{fig:evol3cases}B,C,D). For about 50\% of its lifetime, the magma ocean has a depth equal to or smaller than 50 km for the Ref-A case. The intersection of the adiabat with the rheology front at the two pressure depths where switches in the parameterization of the melting curves occur (see Section \ref{section:meltingCurves} and Appendix \ref{appendix:meltCurves}), results in equal characteristic jumps in MO evolution. 
The decrease in the cooling flux is independent of the decrease in depth $D$, since $D$ is explicitly overwritten when using the soft-turbulence parameterization (see Eqs. \ref{eq:Ra} and \ref{eq:Fsoft}). Nevertheless, $D$ defines the average viscosity within the convecting domain, which enters the $Ra$ calculation (Eq. \ref{eq:Ra}). $Ra$ is ultimately responsible for the decrease in heat flux. 



The role of radioactive decay as energy source in the MO evolution of an Earth-sized planet is insignificant (Fig. \ref{fig:evol3cases}C), unless the planet is formed within few Myr, which includes the contribution of the short-lived elements $^{26}$Al and $^{60}$Fe. Theirs becomes comparable to the long-lived element contribution after 9 Myr and insignificant to the MO evolution by 7 Myr after CAI formation, in agreement with \citet{Elkins2012} findings. 

Comparing the two atmospheric approaches, we find that the pure water vapor grey approximation underestimates the thermal blanketing in comparison to the lbl model because of the low absorption coefficient used to represent the whole thermal radiation spectra ($k_{0,H_2O}=0.01$). 
The lbl pure H$_2$O model resolves better the steam IR absorption, although it overlooks the role of CO$_2$.

The "bb-grey-lbl atmosphere" comparison captures the decreasing convective fluxes at the last MO time step as in the "bb-grey-spectrally resolved atmosphere" comparison of the \citet{Lebrun2013} model. In our approach this is due to the decrease in temperature difference and in $Ra$ towards the MO end. However, upon evaluation at the last time step before solidification the $Ra$ drop to $10^{10}$ is not seen in their work (where $Ra=10^{14}$=const), likely due to differing average viscosity calculation and spatial resolution.

On a technical note, 
a $ Ra$ overshoot towards lower values is observed near the end of the magma ocean phase (Fig. \ref{fig:evol3cases}D). Consequently, the switch to solid occurs abruptly from $Ra=10^{10}$ to 10$^{12}$, two orders of magnitude higher than the value obtained during convection of the last 1-km-deep liquid-like layer of magma ocean. This is a numerical artifact that correlates with high radial resolution of the model layers ($\approx$ 1 km). Therefore, care should be taken when using convective heat flux parameterizations with high spatial resolution very close to the critical melt fraction, because the rheology becomes more complex at high crystal values. 


\subsection{Outgassing and atmospheric build up}\label{section:evolutionoutgassing}

The assumption of greenhouse gases H$_2$O and CO$_2$ as major species is in accordance with an oxidized MO surface \citep{Hirschmann2012b,Zhang2017} and bulk silicate Earth \citep{Lupu2014}. 

\begin{figure}[htb!]
	\centering
	\includegraphics[width=1.05\columnwidth]{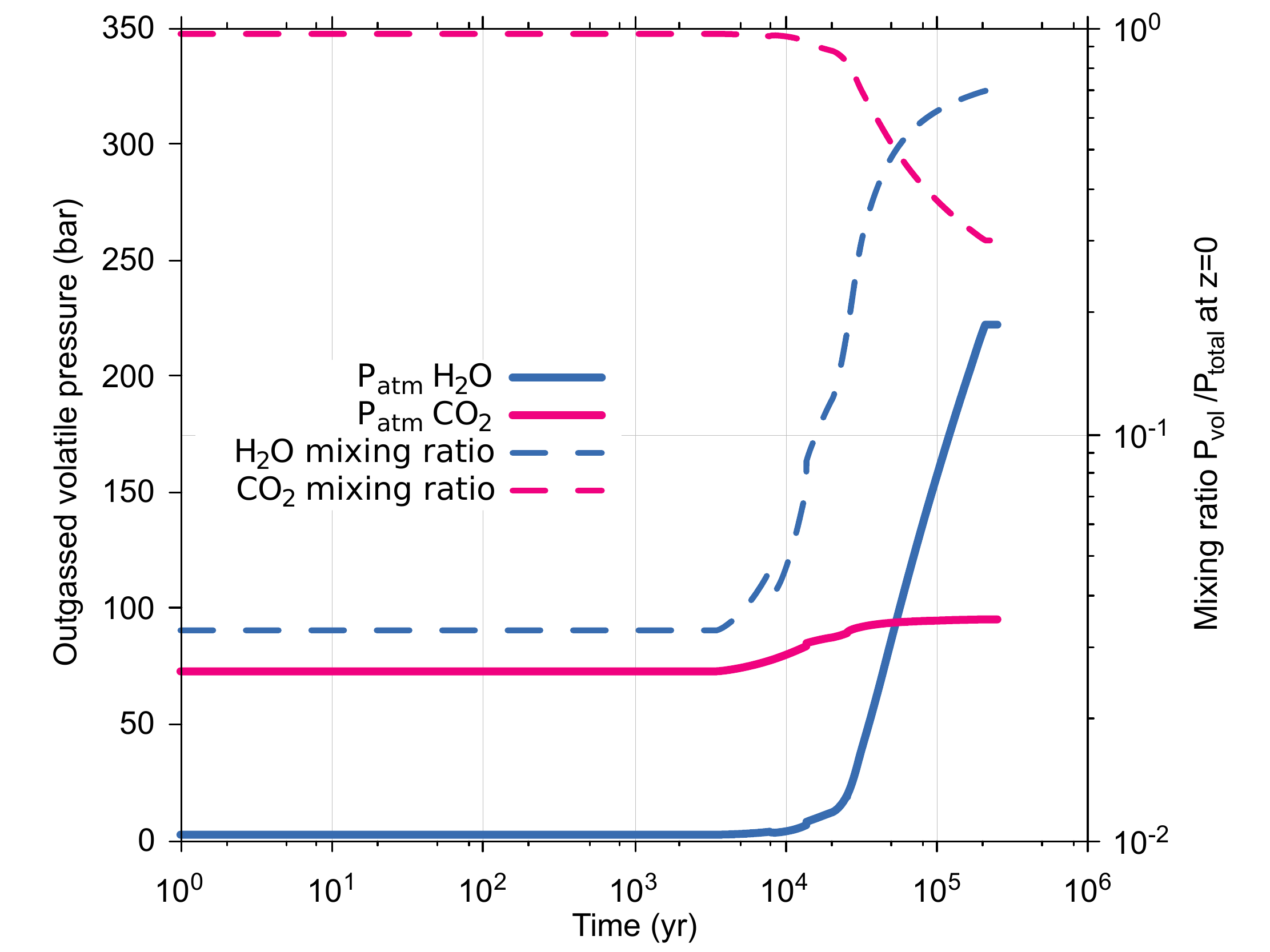}
	\caption{Evolution of H$_2$O and CO$_2$ outgassing based on the Ref-A case (see Table \ref{table:Ref}). Absolute quantity of outgassed volatile in the atmosphere (solid lines) and relative mixing ratio at the surface (dashed lines) are shown.}
	\label{fig:evolOutgassingA}
\end{figure}
\begin{figure}[ht!]
	\centering
	\includegraphics[width=1.05\columnwidth]
	{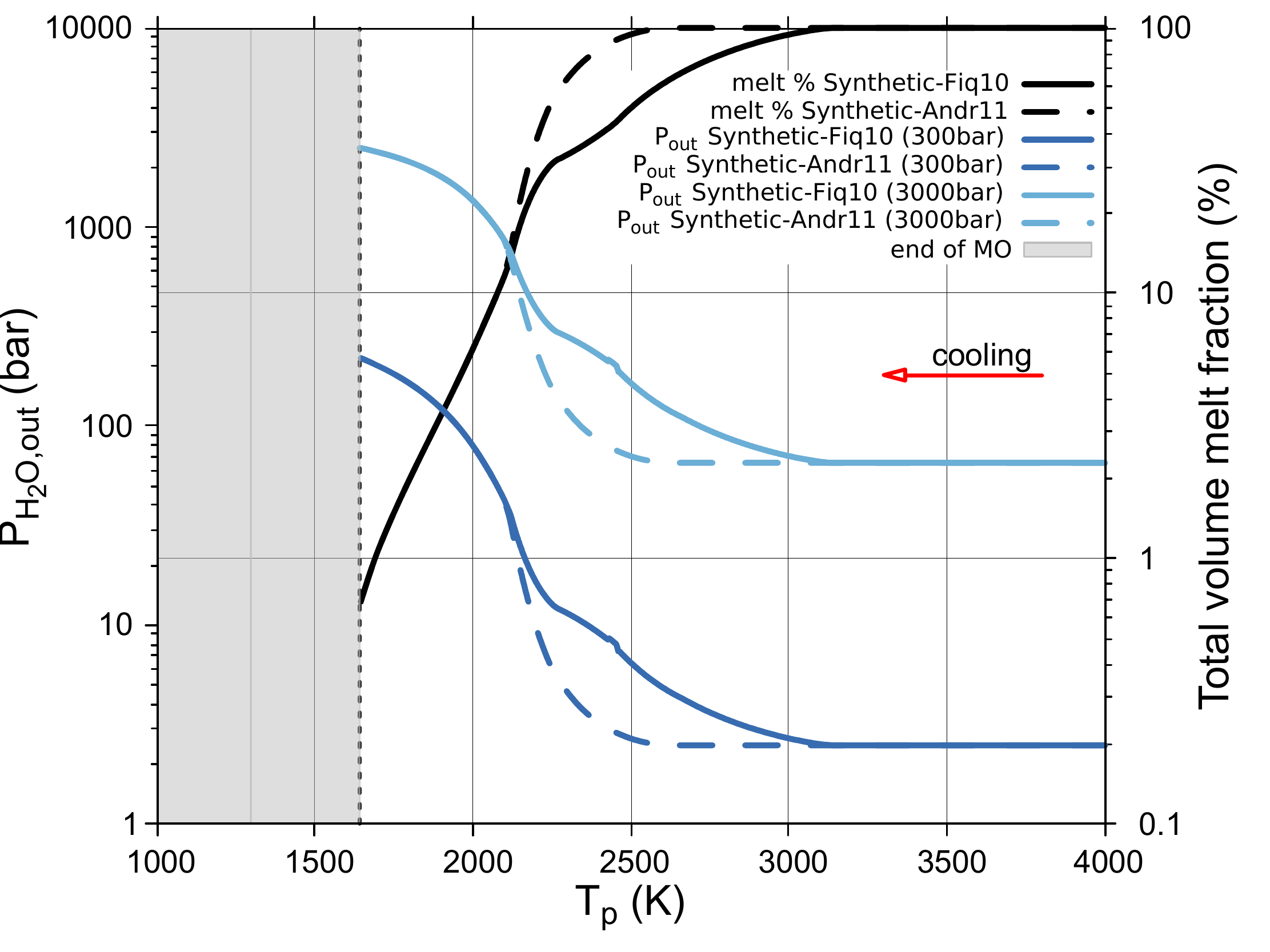}
	\caption{Effect of the choice of melting curves on the variation of the mantle melt fraction with potential temperature. The end of the magma ocean occurs at 99.3\% solidification (grey shaded area). Two sets of melting curves are compared: ``Synthetic-Fiq10'' (dashed lines) and ``Synthetic-Andr11'' (solid lines) that share the \citet{Herzberg2000}, \citet{Hirschmann2000} and \citet{Zhang1994} parameterization for the upper mantle and use the \citet{Fiquet2010} and the \citet{Andrault2011} parameterization for the lower mantle respectively. 
		Different volatile inventories: 300 bar (dark blue lines; Ref-A) and 3000 bar (light-blue) are shown for comparison. The global melt fraction for each case is shown in black color.}
	\label{fig:figAspect}
\end{figure}

As far as the volatile solubility is concerned, molten silicate is a poor CO$_2$ solvent. It thus operates as a ``CO$_2$-pump'' into the atmosphere. In contrast, H$_2$O is highly soluble in the silicate melt and does not leave the mantle until the latest stage of the magma ocean where the enrichment in the melt peaks. The evolving atmospheric composition reflects those features as it transitions from a CO$_2$-dominated to a H$_2$O-rich one (Fig. \ref{fig:evolOutgassingA}).  
The major release (from 2.5 to 220 bar) of the water vapor occurs when the total melt fraction of the mantle reduces from 30\% to 2\% or as the potential temperature drops from $\sim2200$ to $\sim1650$ K (Fig. \ref{fig:figAspect}). This effect is the basis for the so-called ``catastrophic'' outgassing of a steam atmosphere \citep[e.g.][]{Lammer2015}. It reflects the progressive replacement of melt volume with solid volume that has a small capacity for storing volatiles. 

In addition, the choice of melting curves defines the degree of melting throughout the magma ocean lifetime, and similarly affects the accompanying outgassing process. We find that over the range $T_p\in[3000,2200]$ K the melt fraction differs by 10--43$\%$ at the same potential temperature, comparing chondritic and peridotitic composition for the lower mantle (Fig. \ref{fig:figAspect}). The choice of lower mantle melting curves does not affect the final outgassing but modifies the onset of catastrophic outgassing by maximum 5\% of the total volatile volume. Therefore, chondritic composition for the lower mantle disfavors early water release for a cooling magma ocean for potential temperatures above 2200 K.

\begin{figure}[hb!]
	\centering
	\includegraphics[width=1.0\columnwidth]{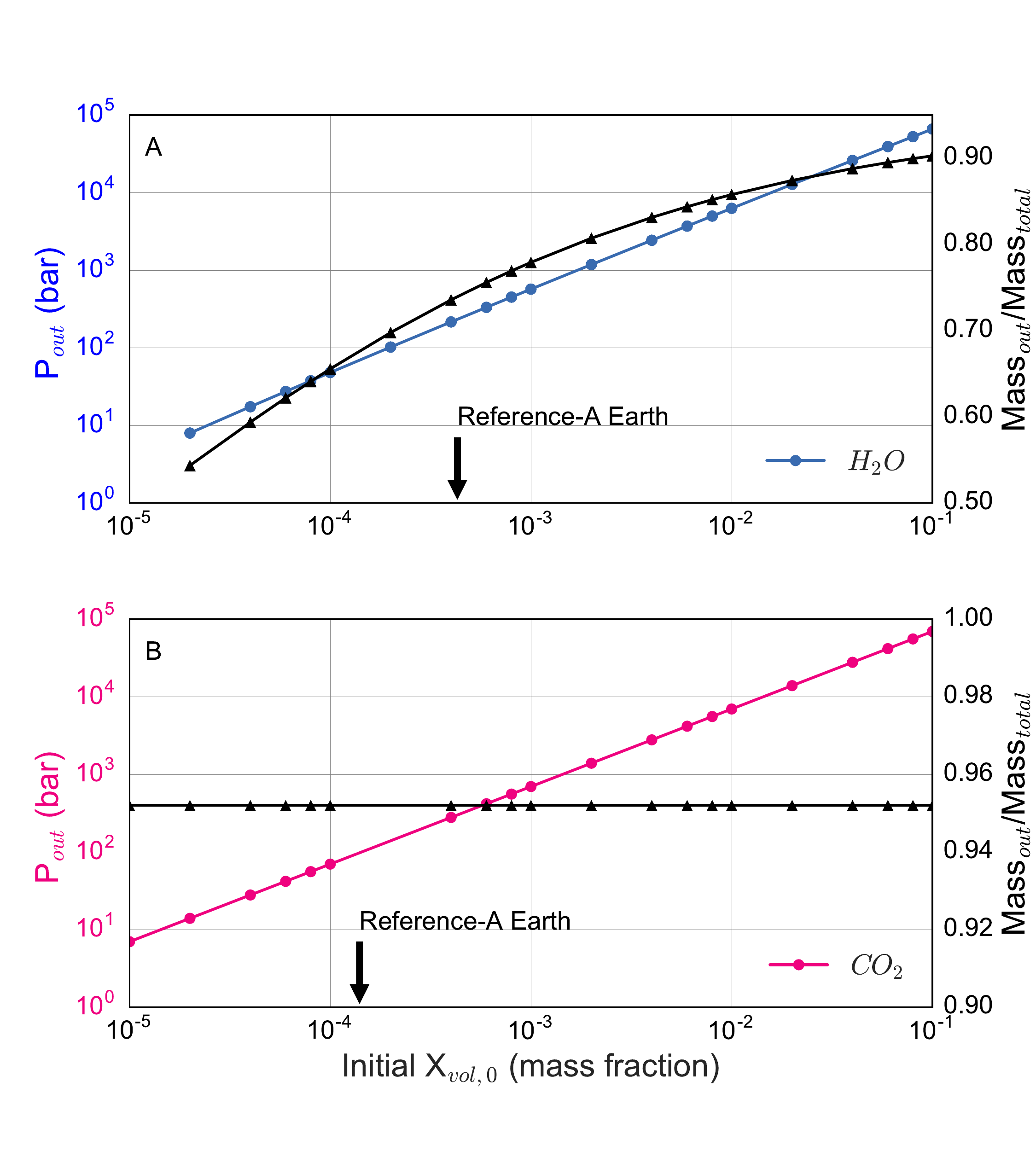}
	\caption{Estimates of maximum outgassing at the end of the magma ocean (99.3$\%$ solid), depending on initial bulk abundance for volatiles H$_2$O and CO$_2$. \textbf{A:} The absolute amount of H$_2$O outgassed by the end of the magma ocean (colored line, left y-axis) is plotted against the initial concentrations in the mantle. The mass of outgassed volatile relative to the mass of the total volatile reservoir is plotted on the right axis (black line, right y-axis). \textbf{B:} Same as in panel A but for the CO$_2$ volatile. The performed experiments are plotted with points.}
	\label{fig:doubleoutgass}
\end{figure}
Simultaneous to the final outgassed quantity, we also calculate the relative volatile inventory extracted from the mantle  assuming different initial concentrations (Fig. \ref{fig:doubleoutgass}). As expected, the higher initial concentration results in higher outgassing. 
However, the relative quantity varies as follows: We find that $[45\%,10\%]$ of the initial water reservoir remains in the mantle for the examined range $X\in[10^{-5}, 10^{-1}]$ respectively, while the rest $[55\%, 90\%]$ is in the atmosphere. This suggests that the lower the initial mantle abundance, the larger is the relative amount of water stored in the planet's interior after the magma ocean ends. By contrast, only $\approx 6\%$ of CO$_2$ remains in the mantle for an Earth--sized planet independently of the initial concentration assumed.

\begin{figure}[htb!]
	\hspace{-1cm} 
	\includegraphics[width=0.55\textwidth]{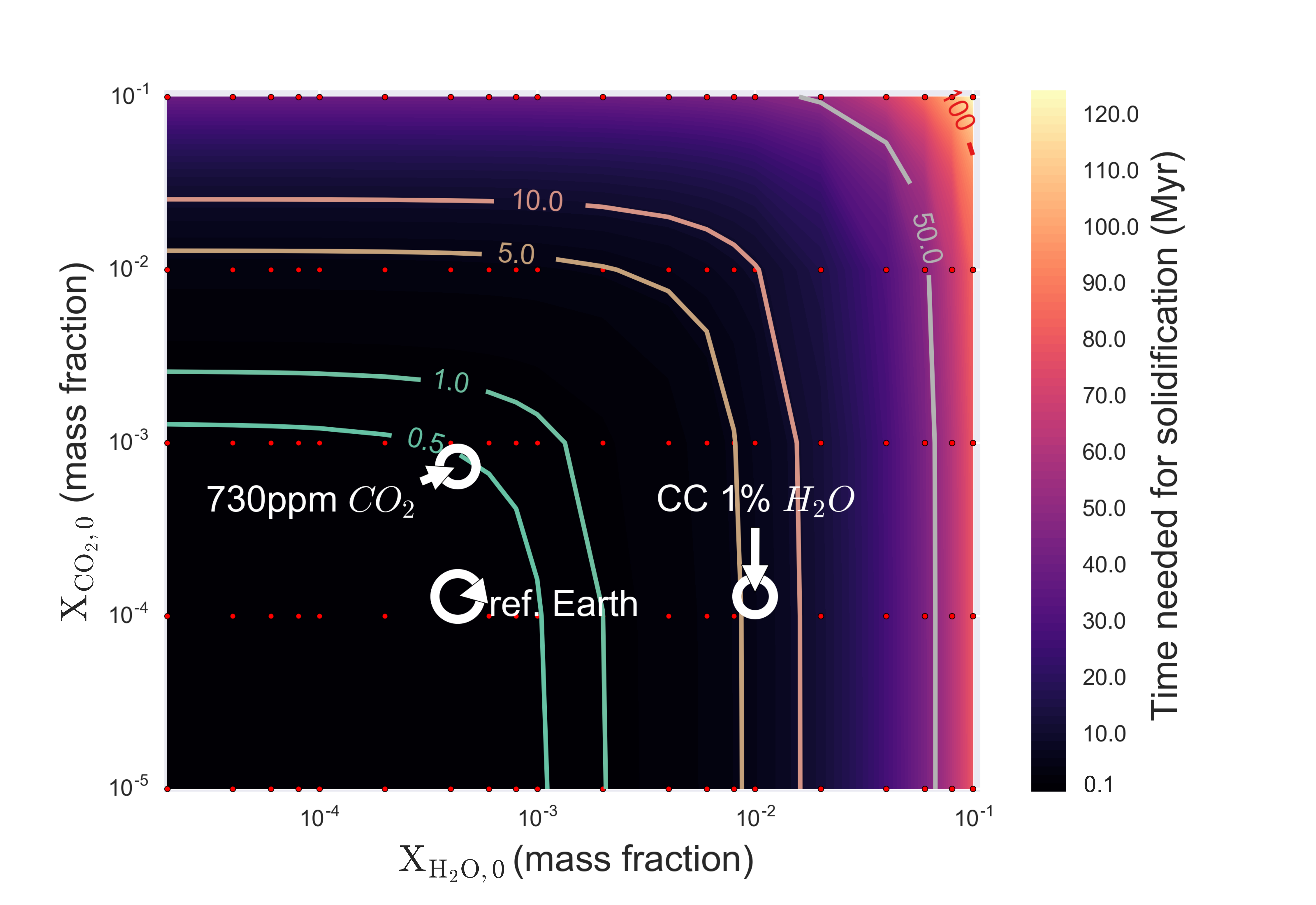}
	\caption{Colormap of minimum solidification time for various initial H$_2$O and CO$_2$ abundances in the mantle, expressed in initial concentrations $X_{volatile,0}$ (at model time 0). \textbf{Open circles} annotate: CC 1 wt\%  H$_2$O-abundance, estimated terrestrial CO$_2$ abundance 730 ppm by \citet{Marchi2016}, and the abundances used in the Ref-A scenario. 
		Red points correspond to the model experiments carried out. \textbf{Isolines} of $t_s$: 0.5, 1, 5, 10, 50 and 100 Myr are plotted for reference.}
	\label{fig:evolOutgassingB}
\end{figure}

\subsection{Effects of model parameters on the MO lifetime}\label{section:sensitivity}

The combined H$_2$O/CO$_2$ inventory was found to delay the MO termination in prior works \citep[e.g.][]{Zahnle1988,Abe97,Lebrun2013}. The \citet{Elkins-Tanton2008} work considers different MO depths (2000km, 1000km, 500km) than our global MO for Earth (2890 km). Consequently, the volatile masses differ for the same assumed concentration and a direct comparison is not possible. Recently, \citet{Salvador2017} have studied the effect of water abundances on the global MO solidification time yielding longer durations likely due to the use of a non-grey atmospheric model. In our study we quantify the solidification time ($t_s$) by sampling a larger domain of initial abundances for the two species and assuming a grey atmosphere (Fig. \ref{fig:evolOutgassingB}). The MO duration amounts to $\approx$ 0.21 Myr for conservative Earth volatile abundances 
while it would reach 5-10 Myr for an (unlikely) Earth-sized planet made entirely out of carbonaceous chondritic (CC) material with 1 wt\% of H$_2$O. Our results confirm that the atmosphere is the most important solidification delaying factor.  



 
However the effect of each separate interior process on the duration of the magma ocean stage remains difficult to disentangle and it would help clarify future modeling priorities. In Table \ref{table:sensitivity} we present an overview of the effect of additional factors and parameters on the MO solidification time ($t_s$). Each $t_s$ is obtained through varying parameters and/or including a different process (first column). The second column states the number of parameters (three at most) that have been modified in each experiment with respect to a reference case. The third column gives the details on the experiment changes with respect to the reference case. We calculate the solidification time ($t_s$, fourth column), as well as the absolute ($\Delta t_s$, fifth column) and relative difference ($\Delta t_s/t_{s,ref}$, sixth column) with respect to the solidification times ($t_{s,ref}$) obtained during two reference simulations: Ref-A (Table \ref{table:Ref}) and Ref-B. The latter uses the same parameters as the Ref-A settings but does not include CO$_2$. We thus obtain the tendency of each factor to increase or decrease the solidification time (``+'' or ``$-$'' sign respectively) as well as its magnitude. 
Below we discuss only the most crucial contributions.

\begin{table*}
\centering
\caption{Overview of the effects of various parameters on the solidification time. Different scenarios are compared to a reference case. The scenarios consist of varying or replacing a parameter or physical process as indicated in the first column. The total number of changed parameters (three at most) with respect to the reference scenario is indicated in the second column. The employed parameter values and/or the description of the process are in the third column. The fourth column shows the solidification time $t_s$, and the fifth and sixth columns the absolute and relative difference of $t_s$ with respect to the reference cases A or B.}
\renewcommand{\arraystretch}{0.8} 
\begin{tabular}{llllll}
 \hline
 \hline
 Modified parameter & $\#$ & Value/ Description &  $t_{s}$ (yr) & $\Delta t_{s}$ (yr) & $\Delta t_{s}/t_{s,ref}$ \\
 \hline
  \textcolor{white}{......}\textbf{Reference-A} & -- & As in Table \ref{table:Ref} & \textbf{208,600} & -- & -- \\
   1a:\textcolor{white}{...}H$_2$O content& 1 & $X_{H_2O,0}=10$ ppm & 58,900 &$-149,700$ & $-72\%$\\
    1b: & 1 &$X_{H_2O,0}=10^5$ ppm & 69,699,000 &+69,490,400& $+33000\%$ \\
      \hline
   2a:\textcolor{white}{...}CO$_2$ content & 1 & $X_{CO_2,0}=10$ ppm & 160,500 &$-48,100$& $-23\%$\\
   2b:  & 1 &$X_{CO_2,0}=10^5$ ppm & 3,919,000 &+3,710,400 & $+18000\%$\\ 
      \hline
   3a:\textcolor{white}{...}Liquid viscosity &1 & $\eta_{l}=f\left(T,X_{H_2O}\right)$ & 213,400 &+4,800$$& $+2\%$\\
    3b:  & 2 & $\eta_{l}=f\left(T,X_{H_2O}\right)$, $X_{H_2O,0}=10^5$ ppm & 69,711,000 &+69,502,400& $+33000\%$ \\ 
    \hline
   4a:\textcolor{white}{...}Radioactive sources &1 & $t_{planet}=100$ Myr & 208,600 & 0$$ & \textcolor{white}{+}$0\%$ \\
    4b:   &2 & $t_{planet}=2$ Myr & 6,036,780 & +5,828,180 & $+2793\%$ \\
        \hline
   5:\textcolor{white}{.....}Heat flux parametrization &1 & $F_{hard};\frac{L}{D}_{max}=1$ & 260,770 & +52,170$$ & $+25\%$ \\ 
    \hline
   6a: \textcolor{white}{..}Upper mantle solidus &1 & ${T_{sol}}-20$ K $\Rightarrow T_{RF,0}=1625 K$ & 216,400 & +7,800$$ & $+4\%$ \\
   6b:  &1 & ${T_{sol}}-50$ K $\Rightarrow T_{RF,0}=1595$ K & 228,700 & +20,100$$& $+10\%$\\
   6c:   & 1 &${T_{sol}}-100$ K $\Rightarrow T_{RF,0}=1545$ K & 250,600 & +42,000$$ & $+20\%$\\
   6d:   & 1 &${T_{sol}}-400$ K $\Rightarrow T_{RF,0}=1245$ K & 434,600 & +226,000$$ & $+108\%$\\
       \hline
   7:\textcolor{white}{.....}Lower mantle melting curves & 2 &$T_{sol,liq}$; \citep{Andrault2011} & 207,100 &-1,500$$ & $-1\%$\\
    \hline
   8:\textcolor{white}{.....}Alternative melting curves &2 & $T_{sol,liq}$; Linear (\ref{appendix:meltCurves}) $\Rightarrow T_{RF,0}=1360$ K & 126,670 &-81,930$$ & $-39\%$\\
    \hline
    9:\textcolor{white}{.....}Irradiation & 1 &$72\%S_0$ & 208,500 &-100$$&  \textcolor{white}{+}$0\%$\\  
   10a: Irradiation \& albedo &2 & $S_0$, $\alpha=0.15$ & 208,600 & 0$$ &  \textcolor{white}{+}$0\%$ \\
   10b: & 2 &$72\%S_0$, $\alpha=0.60$ & 208,500 & -100$$ &  \textcolor{white}{+}$0\%$ \\
      \hline
   11:\textcolor{white}{...}No atmosphere \& $\eta_{l}\left(T\right)$ & 1 &No atmosphere & 2,000 & -206,600$$ & $-99\%$ \\ 
   12a: No atmosphere \& $\eta_{l}\left(T,X_{H_2O}\right)$ & 2 & No atmosphere, $X_{H_2O,0}$=$410$ ppm & 2,958 &-205,642$$& $-99\%$ \\ 
   12b: No atmosphere \& $\eta_{l}\left(T,X_{H_2O}\right)$ & 3 & No atmosphere,  $X_{H_2O,0}$=$10^4$ ppm  & 2,713 & -205,887$$ & $-99\%$ \\
 \hline 
  \\
   \textcolor{white}{.......}\textbf{Reference-B} & -- & As Reference-A with $X_{CO_2,0}=0$ & \textbf{156,700} & -- & -- \\
    13:\textcolor{white}{...}Lbl atmosphere & 1 & Steam lbl & 736,100 & +579,400 & $+278\%$ \\  
 \hline
\end{tabular}
\label{table:sensitivity}
\end{table*}

When accounting for the water dependence of the melt viscosity in exp. 3 we expect a shorter solidification time, that reflects the more efficient convection due to lower viscosity.  $\eta_l$ decreases due to progressive enrichment of water concentration in the melt during the MO evolution (from 410 to $\approx 10^4$ ppm (Ref-A)). The atmospheric radiative forcing remains identical to the Ref-A case. 
The expected cooling acceleration is counteracted by the delaying role of the outgassed vapor atmosphere (Exp. 3), even so for particularly water-rich settings (as seen by the almost identical $t_s$ of water-rich exp. 1b and 3b).
The effect of viscosity on $t_s$ 
becomes evident in the black body cases (Exp. 11, 12). 
With respect to the black body case of experiment 11 ($t_s= 2000$ yr) that uses constant 10 wt\% water content (\citet{Karki2010}), we observe an increase in the solidification time ($t_s= 2713$ yr) in exp. 12b that uses water dependent viscosity. This is explained by the fact that in exp. 12b. the 10\% water enrichment occurs only at the latest MO stage and not throughout the whole run. 
Our parameterizations show that one order of magnitude enrichment in H$_2$O in the melt causes a decrease of up to two orders of magnitude in the viscosity (Fig. \ref{fig:viscosityVFT}). This becomes important at lower melting temperatures $T_{RF,0}< 1400$ K which correspond to evolved silicate melts \citep{Parfitt2008}. Experiments 12a and 12b confirm the tendency we hypothesized for the viscosity role in decreasing $t_s$ with increasing water content (410ppm and 10$^4$ ppm accordingly). Therefore the water-enriched melt accelerates the solidification process and it should be taken into account for evolved surface compositions or planets around EUV and XUV active host stars that lose their atmospheres. According to \citet{Abe97} low viscosity enhances the differentiation of minerals. Therefore, such a $\eta_l$ parameterization is also vital in better modeling the mineral solidification sequence. 

Using the hard turbulence approximation for the convective flux rather than the soft approximation yields a slight increase in the solidification time (experiment 5). The abrupt decrease of $\approx$ 1000 K in the surface temperature at the MO termination is reduced by up to 300 K by employing the hard turbulence parameterization. During this, the $Pr$ number is updated according to the evolution of the liquid viscosity and the flow aspect ratio ($\lambda$) takes values between 1 and 2. 
Significant work that has been done in this direction shows numerical proof of the hard turbulence regime \citep{GrossmannLohse2011} and suggests that it could affect the thermal transport controlled by the boundary layers \citep{Grossmann2003geometry,Lohse2003numericalUtraConv}.

In experiment 6 we examine the role of uncertainty in the upper mantle ($0-22.5$ GPa) solidus. The $\pm 20$ K error estimated in the solidus expression of \citet{Herzberg2000} has a measurable impact $(+4\%)$ on the solidification time. The mere uncertainty in the experimental data can thus affect the magma ocean solidification time by a few thousands of years. 

Further decreasing the upper mantle solidus by 50, 100 and 400 K causes the solidification time to decrease by 10, 20 and 108\% respectively. Compositions more silicate--evolved compared to the KLB-1 peridotite have such lower melting temperatures. The -400 K value corresponds to rhyolite \citep{Parfitt2008}. \citeauthor{Lebrun2013} and \citet{Salvador2017} previously acknowledged that the chemical composition of the magma ocean at its latest stages would be a decisive factor in the evolution. \citet{Schaefer2016} and  \citet{Wordsworth2018} further resolved the chemical evolution for specific compositions. Our result emphasizes the controlling role of the surface melting temperature in the solidification duration and reveals a linear dependence between them. 


The solidification time is however insensitive to changes in the lower mantle melting curves (experiment 7) as long as bottom-up solidification is ensured. The reason is that they affect neither the amount of CO$_2$ in the atmosphere, the majority of which is degassed at the beginning of the magma ocean phase, nor the water enrichment which does not occur at high MO depths.
In experiment 8 we test the effect of linearizing the melting curves of \citet{Abe97}, where the solidification time decreases significantly (-39\%). The higher melt fraction preserved at the end of the magma ocean is tied to lower final outgassing, which explains the difference to the Ref-A setting. \citet{Lebrun2013} has previously discussed a similar effect of the curve linearisation. A quantitative comparison is however inconclusive due to the different atmospheres used. 
\subsection{Qualitative difference between grey and lbl atmospheric blanketing}\label{section:greyVSlbl}

\begin{figure*}[htbp!]
\centering
\includegraphics[width=1.0\textwidth]{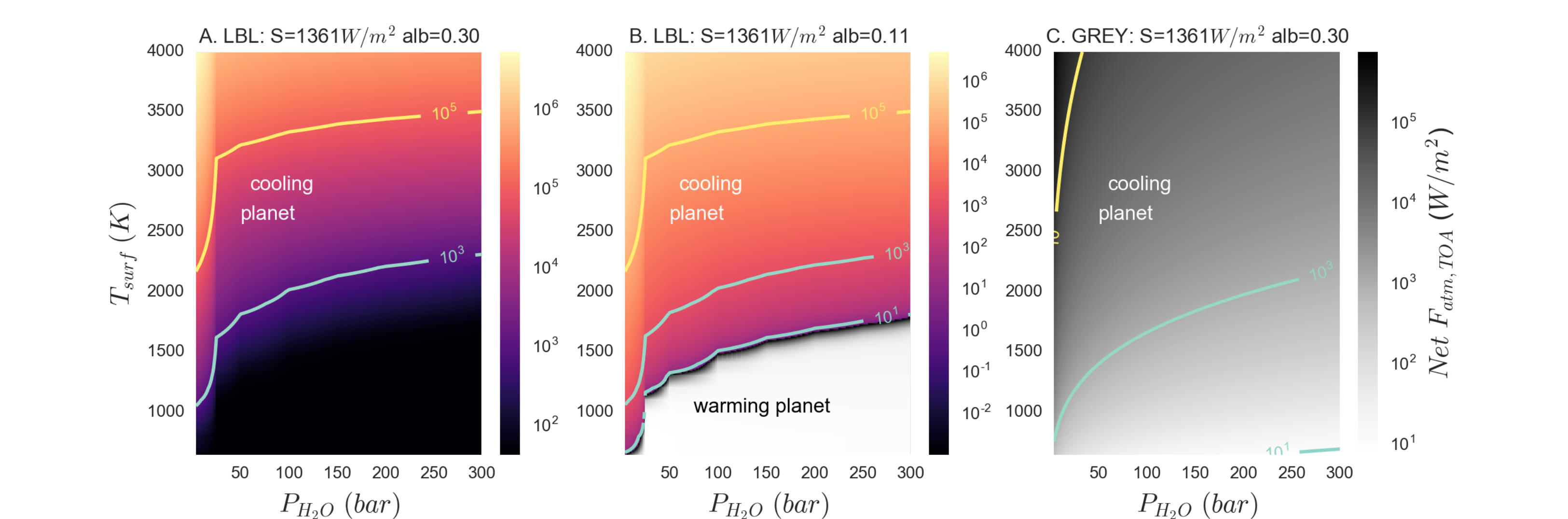}
\caption{Net outgoing radiation flux at TOA for $(P_{H_2O},T_{surf})$ calculated for two specific incoming solar radiations ($F_{Sun}(S=1361 \;\text{W/m}^2, \alpha=0.11) = 303 \;\text{W/m}^2$ and $F_{Sun}(S=1361 \;\text{W/m}^2, \alpha=0.30) = 238\text{ W/m}^2$) employing \textbf{A, B:} the lbl model of \citet{Katyal2017} and \textbf{C:} the grey approximation of \citet{AbeMatsui85} as used in \citet{Elkins-Tanton2008}. In all three plots only the net cooling $F_{atm,TOA}$ (positive sign convention) is shown in the colored legend. The grey approximation results exclusively in cooling fluxes for both cases examined.}
\label{fig:greyVSLBL}
\end{figure*}

We clarify a fundamental difference between the atmospheric approximations that were implemented in this work. We illustrate this by assuming a high ($F_{Sun}(S=1361\; \text{W/m}^2, \alpha=0.11) = 303 \;\text{W/m}^2$) and a low ($F_{Sun}(S=1361 \; \text{W/m}^2, \alpha=0.30) = 238\; \text{W/m}^2$) incoming solar radiation (Fig. \ref{fig:greyVSLBL}). The difference is only in the assumed albedo value, 0.11 or 0.30.

In the lbl approximation (Fig. \ref{fig:greyVSLBL}A, \ref{fig:greyVSLBL}B), the colormap combinations of $P_{H_2O}$ and $T_{surf}$ lead to planetary cooling.
In the high $F_{Sun}$ case, for each value of the surface temperature $T_{surf}$ between 700 and $\approx$ 1700 K there exists a threshold value of outgassed water $P_{H_2O}$ across which the net radiation balance at TOA is negative and the planet warms. This effect is absent in the low $F_{Sun}$ case, which yields a cooling regime for all combinations of $P_{H_2O}$ and $T_{surf}$. 
On the contrary the grey approximation shows a negligible difference of the magma ocean cooling flux of the order of 10$^{-1}$ W/m$^2$, accounting for the $T_{eq}$ of our solar system's inner planet orbits (Fig. \ref{fig:greyVSLBL}C). In fact the grey atmosphere is insensitive to variations in the incoming stellar radiation. 

The reason is that in the grey energy balance (Eq. \ref{eq:Fgrey}), the incoming solar flux enters only in the calculation of the equilibrium temperature. The latter does not vary more than a factor of 2 over the insolation range in our solar system history ($T_{eq}=144$ K for the case of the young Sun and $T_{eq}=256$ K for today's Sun at 1 AU). The fourth power of $T_{eq}$ has a minor contribution compared to the fourth power of the surface temperature of the magma ocean, which is higher than $T_{RF,0}=1645$ K (Ref-A) throughout the evolution.



In the limit of our convecting magma ocean model, we only explore cooling regimes and obtain the relevant solidification times. The convective cooling flux out of the magma ocean $F_{conv}$ requires $T_{surf}<T_p$ to ensure the necessary gravitational instability for convection to occur (see Eq. \eqref{eq:Fsoft}). 
However, if the flux at the TOA becomes negative (RHS of Eq. \eqref{eq:FLBLBalance}) the system would warm resulting in $T_{surf} > T_p$, a condition which describes a stably stratified system that will not convect. 

The remaining three Sections \ref{section:mechanism}--\ref{section:multiAlbedo} focus on the cooling/warming limit found with the lbl atmosphere.

\subsection{Lbl atmosphere: separating continuous from transient magma oceans}\label{section:mechanism}
%

The lifetime of a magma ocean with a steam atmosphere is controlled by the longwave radiation through its steam layer, the energy received from the star, and the melting temperature of the mantle at its surface. All above factors combine into a comprehensive mechanism that distinguishes between a ``transient'' (or ``short-term'', or ``type-I'' after \citet{Hamano2013}) and ``continuous'' (or ``long-term'', or ``type-II'' after \citet{Hamano2013}) MO evolution path. \citet{Goldblatt2015_waterworlds} and \citet{Ikoma2018} have discussed the warming/cooling distinction, always in relation to the constant radiation limit for the runaway greenhouse (RG) $\approx$ 300 W/m$^2$. We exemplify this idea with an emphasis on the additional role of $T_{RF,0}$. 

\begin{figure*}[htb!]
\centering
\includegraphics[scale=0.57]{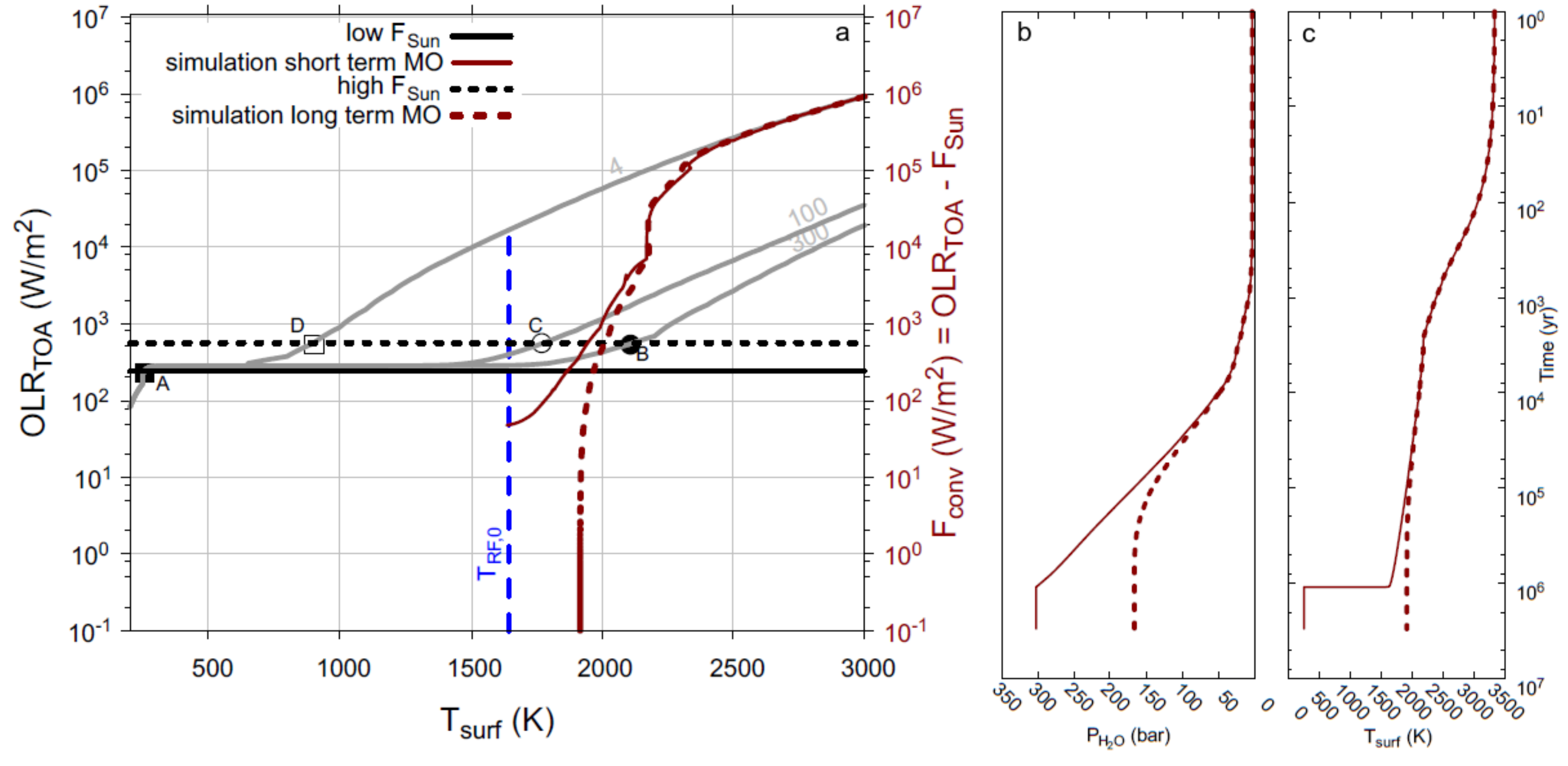}
\caption{\textbf{a:} Mechanism for separating a continuous (long-term) from a transient (short-term) magma ocean as a function of surface rheology front temperature $T_{RF,0}$ \textbf{(dashed blue line)}, isovolatiles of outgassed water pressure at the surface (grey lines), and incoming solar energy $F_{Sun}$ \textbf{(solid and dashed black lines)}. 
Two experiments with low and high $F_{Sun}$ are performed using the same total water reservoir (405 bar). A short-term \textbf{(red solid line)} and a long-term evolutionary case \textbf{(red dashed line)} assuming respectively low and high insolation conditions (see text for values) is shown. Solar insolation is read on the left y-axis. The evolution of $F_{conv}$ is read on the right y-axis. Points A, B, C and D mark the intersection of the isovolatile curves with the value of $F_{Sun}$ considered and are used to explain different evolution scenarios. Isovolatiles cover surface pressures within 4--300 bar. \textbf{b:} $P_{H_2O}(t)$ and \textbf{c:} $T_{Surf}(t)$ evolution for short-term and long-term MO. All parameter values unless otherwise explicitly mentioned are as in Ref-A.}
\label{fig:Goldblatttype}
\end{figure*}

We use two simulations that are subject to different insolation conditions, namely $F_{Sun,low}=238$ W/m$^2$ and $F_{Sun,high}=563$ W/m$^2$ (Fig. \ref{fig:Goldblatttype}a black solid line and black dashed line respectively), leaving all other parameters unchanged. The $F_{Sun,high}$ is obtained using $S=2648$ W/m$^2$ that corresponds to the incident radiation at the orbital distance of Venus for today's Sun and $\alpha$=0.15, while the $F_{Sun,low}$ is equal to the incoming radiation at Earth orbit today. 
Since $F_{Sun}$ is independent of $T_{surf}$, it is plotted as a line parallel to the $T_{surf}$ axis (Fig. \ref{fig:Goldblatttype}a).  
Both simulations have the same water reservoir (405 bar or 550 ppm initial concentration) to ensure outgassing of 1 Earth ocean (300 bar) at the end of the magma ocean stage. $OLR_{TOA}$ as a function of $T_{surf}$ is plotted for three values of atmospheric water content (4, 100, and 300 bar), which we term ``isovolatiles''  (grey lines).
$F_{Sun}$ intersects with each isovolatile over a temperature value $T^\prime_{surf}$. The cooling flux $F_{conv}$ (read on the right axis) becomes zero for that specific water content and the planet ceases to cool. If $T^\prime_{surf}$ is higher than the mantle rheology front temperature at the surface ($T_{RF,0}$), the steam quantity indicated by the respective isovolatile balances the energy flux from the star and the MO does not solidify.



Firstly, we examine the trajectory of the convective flux of the transient magma ocean on the (T$_{surf}$, $F_{conv}$) plane as it cools from $T_{surf}= 3000$ K and with an insolation $F_{Sun,low}$ (Fig. \ref{fig:Goldblatttype}a, red solid line). 
$F_{conv}$ progressively crosses isovolatiles of higher water content. As it approaches the highest outgassed quantity of 300 bar, the difference $OLR_{TOA} -\text{F}_{Sun,low}=\text{F}_{conv}$ remains always positive since the 300 bar isovolatile allows the system to dispose of heat at a higher rate than it receives solar radiation. The high convective flux value ensures cooling until $T_{surf} = T_{RF,0}$, which marks the end of the magma ocean. 
The abrupt cooling after the end of the magma ocean stage and the final outgassing quantity are shown in the evolution of $T_{surf}$(t) and $P_{H_2O}$(t) (Fig. \ref{fig:Goldblatttype}b, \ref{fig:Goldblatttype}c.) 

Secondly, we obtain a long-term magma ocean (Fig. \ref{fig:Goldblatttype}a, red dashed line) in a scenario that assumes $F_{Sun,high}$. Initially, for high values of $T_{surf}$, the same amount of water as before is outgassed and its $F_{conv}$ almost coincides with the one of the short-term case (the difference hardly noticeable on the logarithmic graph is $\approx 10^3$ W/m$^2$). During evolution the outgassing proceeds and the simulation trajectory crosses isovolatiles of higher water content. $F_{conv}$ drops to very low values that tend to numerical zero for $T_{surf}=T^\prime_{surf}\approx 1915$ K. 
The intersection of the incoming radiation $F_{Sun,high}$ with the respective isovolatile over $T^\prime_{surf}$ reflects the steam atmosphere already outgassed when the system ceased to cool. We obtain a point that falls between the isovolatiles of 100 and 300 bar (167 bar read in Fig. \ref{fig:Goldblatttype}b). Consequently, a continuous magma ocean is maintained at potential temperature $\approx T^\prime_{surf}$ (Fig. \ref{fig:Goldblatttype}c) due to a specific combination of incoming solar radiation, its intersection with the 167 bar isovolatile, and the solidification temperature (Fig. \ref{fig:Goldblatttype}a). Note that the long term MO ocean is maintained with less water than one Earth ocean and at an insolation higher than the RG limit. 


The prominent role of $T_{RF,0}$ on the MO type becomes evident when comparing the point ($T^\prime_{surf}$, $P_{H_2O}$) where the isovolatiles intersect $F_{Sun}$, with $T_{RF,0}$. For the short-term magma ocean the intersection point A occurs well below $T_{RF,0}$. That magma ocean stage will be transient for every possible outgassing scenario within the [4,300] bar range. In the case of the higher solar irradiation, we have intersection points with each isovolatile (B, C, and D), which indicate different thermal evolution paths. On the one hand, the points B and C are located at surface temperatures higher than $T_{RF,0}$, which means that if the MO has outgassed the respective quantities of 300 and 100 bar by the time $T^\prime_{surf}$ is reached, it will cease cooling. 
On the other hand, the point D corresponds to a much lower temperature than $T_{RF,0}$, which means that a steam atmosphere of 4 bars under those insolation conditions can counteract the cooling process only if $T_{surf}$ decreases to 900 K. The respective magma ocean stage is transient, since it solidifies at a much higher temperature (i.e. 1645 K). The variation of the OLR as a function of P,T is explored in detail in the companion work. 


\subsection{Lbl atmosphere: Role of orbital distance and albedo on MO evolution} \label{section:orangeAlbedo}

Clearly, the essential quantity regarding the planetary heat budget is $F_{Sun}$, since it distinguishes the fate of the magma ocean between transient and continuous. Below we refer to this limiting incoming flux as $F_{lim}$ (where $F_{conv}$=0) and we specify the incident solar radiation and albedo combinations which satisfy it. 

On combining the stellar luminosity ($L_{Star}$):
\begin{equation}
 	L_{Star}=\left(4\pi R_{Star}\right)^2 \sigma T^4_{eff,Star},
\label{eq:stellarLuminosity}
\end{equation}
where $R_{Star}$ is the stellar radius and $T_{eff,Star}$ the effective temperature at the star's photosphere; with the expression of \citet{Gough1981} for the evolution of solar luminosity we get $T_{eff,Star}(\tau)$. Combining with the blackbody radiation law for the equilibrium temperature of a planet, we obtain the following equation:
\begin{equation}
 	R=\frac{R_{Star}\cdot T_{eff,Star}^2(\tau)\cdot \sqrt{1-\alpha_{max}}}{2 \sqrt{\displaystyle\frac{F_{lim}(P_{H_2O},T_{RF,0})}{\sigma}}}.
\label{eq:albedoMax}
\end{equation}
Eq. \eqref{eq:albedoMax} relates to the $F_{lim}$ the maximum albedo $\alpha_{max}$ that an Earth-sized planet at orbital distance $R$ from a star of effective temperature $T_{eff,Star}$ can possess in order to maintain a continuous magma ocean stage. 
The limiting flux included in the denominator of Eq. \eqref{eq:albedoMax} is not constant but equal to:
\begin{equation}
F_{lim}(P_{H_2O},T^\prime_{surf})=\frac{\left(1-\alpha_c\right) S(\tau)_{1AU}}{4},
\label{eq:albedoMax2}
\end{equation}
%
where $\alpha_c$ the critical albedo found with a sensitivity experiment for a given planetary volatile inventory. There $S(\tau)_{1AU}$ is the solar constant at an orbital distance of 1 AU and stellar age $\tau$, $P_{H_2O}$ (in bar) the mass of the water vapor outgassed and $T^\prime_{surf}$ is the temperature over which the stellar insolation crosses the $P_{H_2O}$ isovolatile (Section \ref{section:mechanism}). The obtained limiting flux $F_{lim}$ maps to the data product $OLR_{TOA} (T^\prime_{surf}, P_{H_2O})$. 
Using the \citet{Katyal2017} values of the limiting flux, we compared our Eq. \eqref{eq:albedoMax} with an equivalent expression calculated by \citet{Hamano2013}. The solution is similar with minor differences due to the astrophysical properties assumed. We generalise the formulation to cover our Sun or any other host star with a known photospheric temperature $T_{eff}$ and radius.

\subsection{Lbl atmosphere: dependence of F$_{lim}$ on the melting temperature and steam mass} \label{section:multiAlbedo}

The irradiation conditions which can pinpoint an Earth-sized planet stalling in a magma ocean stage just above the 40\% melting temperature, are extended here to include a range of steam atmosphere masses that span [4,300] bar for two different $T_{RF,0}$ values (the lower melting temperature is representative of a more evolved composition than the KLB-1 peridotite of Ref-A). 

\begin{figure}[htb!]
\centering
\includegraphics[width=1.0\columnwidth]{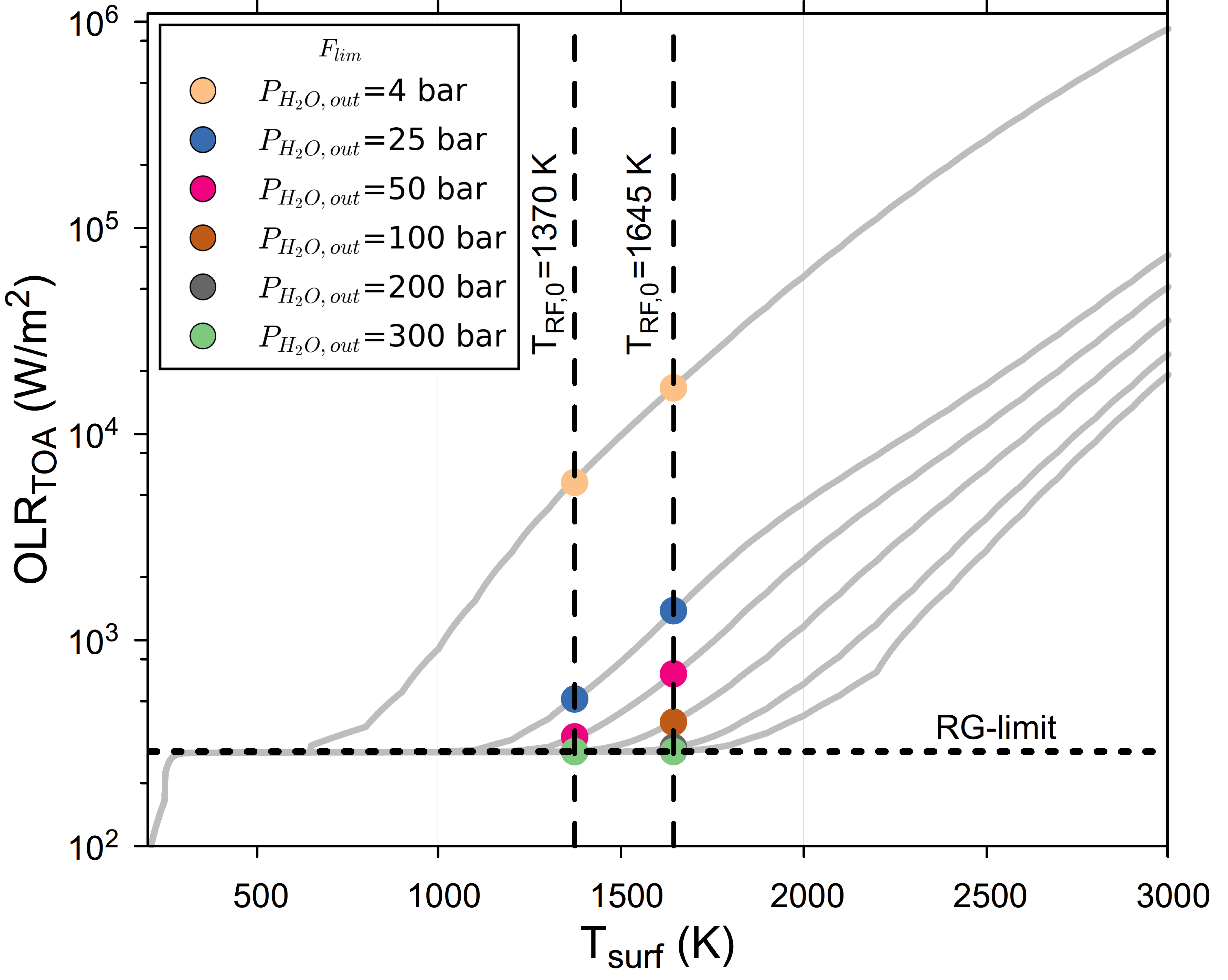}
\caption{OLR$_{TOA}$ as a function of surface temperature and outgassed water surface pressure 4--300 bar, based on data from \citet{Katyal2017}. We consider two different surface solidification temperatures: $T_{RF,0}= 1370$ K as in \citet{Hamano2013} and $T_{RF,0}= 1645$ K as in Ref-A case. Colored points correspond to OLR$_{TOA}$ values obtained for the respective isovolatiles for different outgassed steam atmospheres 4, 25, 50, 100, 200 and 300 bar that overlie magma oceans of different $T_{RF,0}$ (see Table \ref{table:Flimits} for explicit values). Note the variation in temperature coverage of OLR=const=282 W/m$^2$, for different isovolatiles. Data by \citet{Naka92} are used to complement the plot in the region where $T_{surf}\ll T_{H2O,crit}$. }
\label{fig:GoldblattMulti}
\end{figure}

Different radiative flux limits $F_{lim}$ are obtained for the two values $T_{RF,0}=1370$ K and 1645 K, depending on the vapor amount (Fig. \ref{fig:GoldblattMulti}). From the superposition of points that correspond to 200 and 300 bar at $T_{RF,0}= 1645$ K in Fig. \ref{fig:GoldblattMulti}, we note the tendency of steam atmospheres exceeding 200 bar to converge to the constant RG limit (RG=282 W/m$^2$ \citet{Katyal2017}). For $T_{RF,0}= 1370$ K atmospheres already equal to or higher than 100 bar suffice to reach the RG limit. A similar tendency is shown in \citet{Hamano2013}.
 
\begin{figure}[ht!]
\hspace{-0.5cm}
\includegraphics[trim={0 3mm 0 3mm},clip,width=1.1\columnwidth]{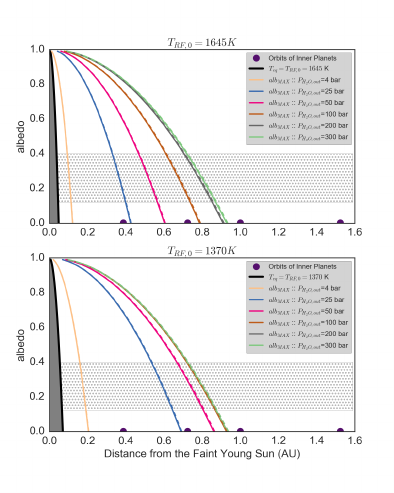}
\caption{Maximum albedo that a planet with a 4--300 bar steam atmosphere can possess at a given distance from the young Sun in order to maintain a long-term magma ocean, assuming two values of $T_{RF,0}$. We plot the critical values that separate continuous from transient magma ocean cases calculated with the use of Eq. \eqref{eq:albedoMax} for several $P_{H_2O}$ (colored lines) and employing the respective steam atmosphere mass limiting outgoing longwave fluxes (extracted from Fig. \ref{fig:GoldblattMulti}; see Table \ref{table:Flimits} for values).
Dashed lines are obtained using the equation of \citet{Hamano2013} employing our $F_{lim}$ limits. 
Not all obtained albedos are realistic. Hatched region shows the possible range of albedos for a cloudless steam atmosphere \citep{Kasting1988, Leconte2013, Goldblatt2013}. Black solid lines mark distances from the star for which $T_{eq}$ is equal to the surface melting temperature of the magma ocean for the full albedo range 
(permanent magma ocean).}
\label{fig:albedoMulti}
\end{figure}

We also find that at lower steam contents the $F_{lim}$ is greater than the RG-limit. All $F_{lim}$ values can be found in Table \ref{table:Flimits}.  
Using Eq. \eqref{eq:albedoMax} we calculate the orbital distance-albedo combinations for which the radiation limits (Table \ref{table:Flimits}) of different isovolatiles are attained (Fig. \ref{fig:albedoMulti}). We assume the solar luminosity at the beginning of its main sequence evolution at $\tau= 100$ Myr (72\% of today's value) \citep{Gough1981}. Not all the calculated albedo values are realistic. The albedo for a cloudless steam atmosphere, based on 1D models and 3D global circulation model calculations lies between [0.15, 0.40] \citep{Kasting1988,Goldblatt2013,Leconte2013,Pluriel2019}. 


Apart from the new radiation limits found, our results are in line with those of previous studies, as far as the insolation role is concerned. In \citet{Hamano2013} the threshold distance between continuous and transient MO types for albedo 0.3 and solar constant 0.72 $S_0$ is 0.77 AU, whereas under the same conditions our calculations show 0.79 AU. The difference is due to the lower absolute OLR steam atmosphere limit of $282 \pm 1$ W/m$^2$ that we obtain compared to the 294 W/m$^2$ limit employed in that study. 

By raising the albedo to the critical value $\alpha_c=0.146$ found in our simulations for an Earth-sized planet at 1 AU that outgasses 1 earth ocean at today's sun, we obtain $\approx$ 12 Myr MO duration. The longer solidification times of 10--30 Myr reported in \citet{Hamano2013} at the same $F_{Sun}$=285.5 W/m$^2$ for the same steam atmosphere are due to the lower surface melting temperature used ($T_{RF,0}$=1370 K).

%

With an albedo of 0.63, a solar constant of 0.7 $S_0$, and total planetary water content X$_{H_2O,0}$= $5.53\cdot 10^{-2}$ wt$\%$ (equivalent to 405 bar) and X$_{CO_2,0}$= $1.4\cdot 10^{-2}$ wt$\%$ (equivalent to 100 bar), \citet{Lebrun2013} found that the distance at which the outgassed water vapor could no longer condense is 0.67 AU. 
Under the same conditions, excluding the influence of CO$_2$, we find in our model that the atmosphere would exist in a runaway greenhouse state trapped in a continuous magma ocean at a critical distance of 0.59 AU. The reason for this discrepancy is two-fold. Firstly, our lbl model approach does not include CO$_2$ that also contributes to the greenhouse effect. 
Secondly, the absolute OLR limit used by \citet{Lebrun2013} is $\approx$200 W/m$^2$ \citep{Marcq2012} (see \citet{Marcq2017} for an updated limit). This is substantially lower than the limit of 282 W/m$^2$ used in our study. Therefore, the shift of our limit inward towards the star corresponds to the higher critical flux that needs to be achieved in order to trigger the qualitative shift from a transient to a continuous magma ocean regime. 
Considering the orbital distances of the inner terrestrial planets of the solar system (Mercury--0.38 AU, Venus--0.72 AU, Earth--1 AU and Mars--1.52 AU), we find that the planets inwards of Earth could sustain a continuous MO within the range of albedos expected for a cloudless steam atmosphere (Fig. \ref{fig:albedoMulti}). Moreover, a 100-Myr-old Earth at 1 AU around the Sun cannot exist in a continuous MO state under any albedo for a steam atmosphere of up to 300 bar (Fig. \ref{fig:albedoMulti}) or of up to 1000 bar according to the recent study of \citet{Ikoma2018}. 

Note that in this work ``continuous'' magma ocean refers to planets that would cease cooling if the amount of steam in the atmosphere was conserved. This cannot be ensured under atmospheric escape processes, which have not been accounted for, and as such the limits calculated here yield the furthest possible distance from the Sun for achieving a continuous MO with a constant atmospheric steam content. 




Using the database of $F_{lim}$ that depends on both the atmospheric water content and $T_{RF,0}$ that we provide in this (Table \ref{table:Flimits}) and in the companion work, Eq. \eqref{eq:albedoMax} is qualitatively extended. It covers MO type transitions for intermediate levels of outgassing below the 300 bar reference value, hence has higher $F_{lim}$. This database is backwards compatible and can also be used in the \citet{Hamano2013} equation.  

\subsection{Evolutionary and permanent magma oceans}\label{section:MOtypes}
We draw specific attention to the difference between ``evolutionary'' and ``permanent'' magma ocean which is studied in other works \citep[e.g.][]{Hammond2017}. Both the transient and continuous magma ocean that we study belong to the so called ``evolutionary'' magma oceans generated during the accretional process. However, within a certain orbital distance the energy for melting the mantle is already provided by the solar irradiation alone and the atmosphere blanketing effect becomes irrelevant. This is the ``permanent'' magma ocean caused by the star. Radioactivity and delivery of kinetic energy cause the evolutionary magma ocean. 
To help distinguish these, the minimum distance from the star for which the equilibrium temperature $T_{eq}$ equals the 40\% melt fraction surface temperature (where permanent MO stage ensues) is indicated by the black lines in Fig. \ref{fig:albedoMulti}. Further research, which is beyond the scope of this study, could lead to an expansion of the orbital distance defining the permanent MO, on accounting for climate feedbacks that raise the surface temperature above the $T_{RF,0}$ melting point.

\subsection{Magma oceans on other planets}\label{section:exoplanets}

\begin{figure*}[htb!]
	\centering
	\includegraphics[width=0.7\textwidth]{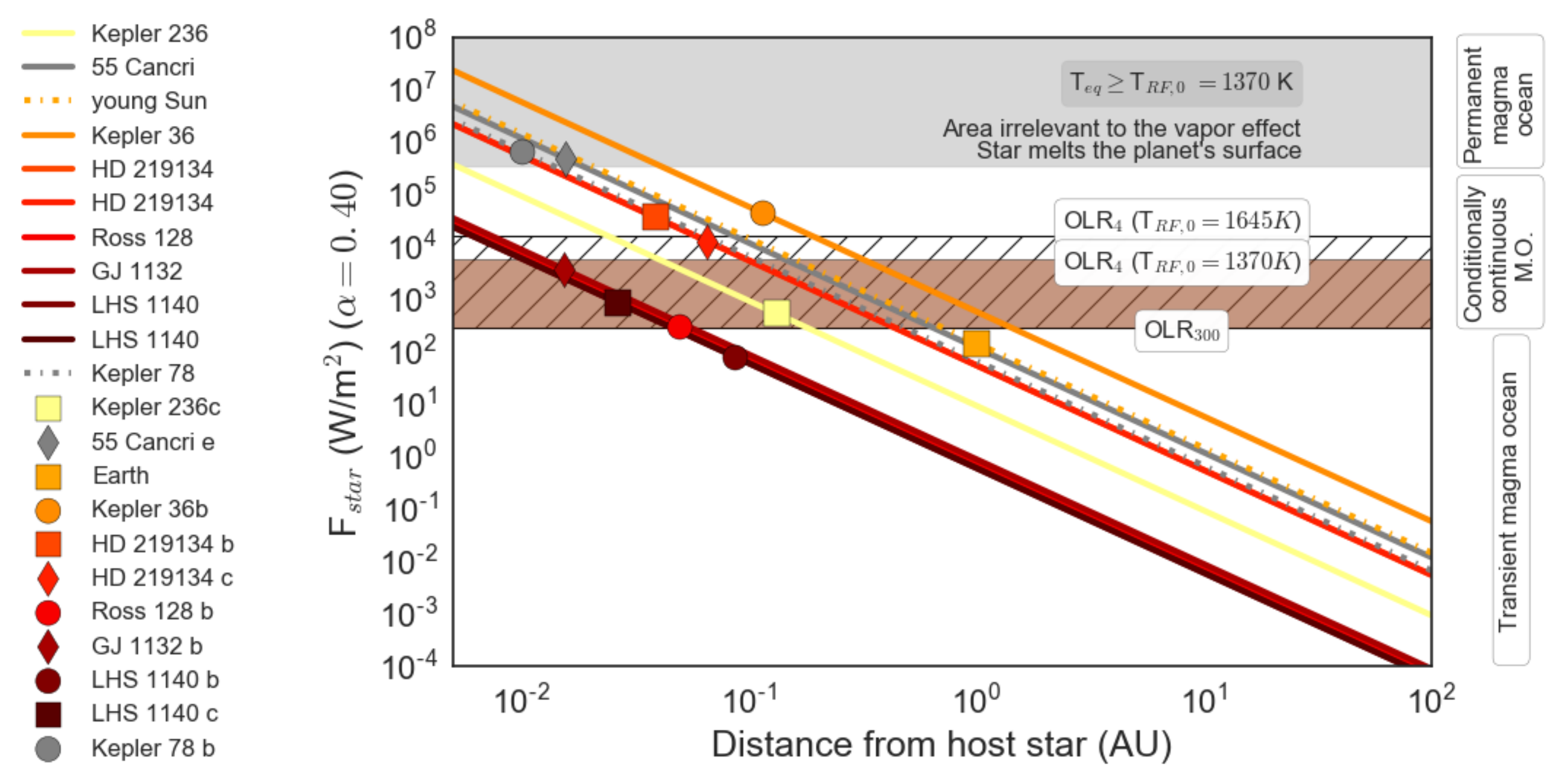} 
	\caption{Incoming stellar energy flux at various orbital distances around M and G stars for a planet with albedo $\alpha$=0.4 (cloudless water vapor maximum value). Examples of potentially rocky exoplanets are plotted on the relevant host star curve. Grey shaded region corresponds to permanent magma ocean at the lowest mantle melting temperature considered. Regions are drawn for 4--300 bar OLR for two different rheology front solidification temperatures: T$_{RF,0}=1370$ (line-hatched) and 1645 K (brown shaded).}
	\label{fig:exoplanet}
\end{figure*}

$F_{lim}$ is not significantly affected by the gravitational acceleration of the planet as long as this has between 0.1 and 2 Earth masses \citep{Goldblatt2013}. For greater planetary mass the pressure levels in the atmosphere change height, as does the level of opacity depth which is crucial for the calculations of the outgoing radiation. \citet{Goldblatt2013} also calculated that for a planet of half the Earth mass, the OLR limiting flux is lowered by only 5 W/m$^2$. In comparison, $F_{lim}$ $\approx 282$ W/m$^2$ $\pm$ 1 W/m$^2$ as calculated for the Earth by \citet{Katyal2017} has a lower uncertainty. Therefore, Eq. \eqref{eq:albedoMax} can be applied  without loss of generality to planets between 0.1 and 2 Earth masses, using the $F_{lim}(P_{H_2O},T_{RF,0})$ (Table \ref{table:Flimits}) calculations by \citet{Katyal2017} .
 
Given their similarity in mass and radius, the criteria for a continuous magma ocean applied for the Earth can be extended to Venus. A continuous magma ocean could not have been possible for the Earth during the young Sun period for any bulk water abundance. 
We find however, that Venus orbit qualifies for a long-term magma ocean within a wide range of planetary albedos [0.15, 0.40] proposed for cloud-free steam atmospheres, as long as its outgassed steam atmosphere amounts to 200 bar or more for a surface solidification temperature of 1645 K (Fig. \ref{fig:albedoMulti}). In the case of the lowest solidification temperature ($T_{RF,0} = 1370$ K), the minimum atmosphere required for a continuous magma ocean at Venus orbit is 50 bar (Fig. \ref{fig:albedoMulti}). This highlights that the melt composition alone could dictate a different magma ocean evolution path for two hypothetical planets with equal water vapor atmosphere masses.

It is additionally important to consider whether a planetary body has had a long impact history or has chemically evolved before impacts remelt it into a magma ocean \citep{Lammer2018}. Such bodies could more easily maintain a secondary continuous magma ocean. Due to their lower $T_{RF,0}$ they would require smaller steam atmospheric mass, instead of the reference one Earth ocean (300 bar) usually assumed in runaway greenhouse studies. 
On the contrary, a chemically unevolved silicate primitive composition that melts at high temperatures would require a massive steam atmosphere $>$100 bar in order to maintain a continuous magma ocean. We conclude that past events of chemical alteration may influence the fate of the magma ocean under the same orbital configuration. Therefore the age of the star and of its planetary system matters. Evolution of the mantle composition during the MO solidification \citep{Elkins-Tanton2008,Schaefer2010} will be an additional factor that prolongs the MO lifetime if it results in decreased $T_{RF,0}$. 

Interestingly, the drop of surface temperature during cooling combined with the tendency of stellar environments to gradually strip planets of their atmospheres \citep{Johnstone2015, Odert2017,Lammer2018} (therefore lowering the surface pressure $P_{H_2O}$) could result in the same outgoing radiation limit during planetary evolution. We see this in Fig. \ref{fig:GoldblattMulti} taking any constant OLR value that crosses multiple isovolatiles. In that case the stellar evolution plays a primary role in the fate of the continuous magma ocean. A G-star with increasing luminosity with time \citep{Gough1981} favors the maintenance of an existing magma ocean because it contributes warming at the critical distance. In contrast, continuous magma oceans will be more elusive around M-stars whose luminosity decreases with time \citep{Baraffe2015}. This is because a continuous magma ocean close to its critical distance will receive less and less stellar radiation eventually creating a window of cooling. A buffer against this effect is the additional vapor outgassing that increases the opacity and lowers the required $F_{lim}$. However, during progressive cooling the interior will exhaust its water supply into the atmosphere. Under these conditions only water--rich planets can sustain a continuous magma ocean. This shows that there are numerous processes that affect MO feasibility. Consider also the possible Trappist-1 exoplanet migration scenarios \citep{Unterborn2018} (suggested in order to justify water-rich composition). 


We explore our findings in view of potentially rocky exoplanets having radius and/or mass within few Earth units (parameters in Appendix \ref{appendix:exoplanets}). Results suggest that there are orbital regions where the magma ocean can be transient, permanent and an intermediate region where it is ``conditionally'' continuous (Fig. \ref{fig:exoplanet}). ``Conditionally'' here refers to the dependence on water content and rheology front temperature. We observe the overlap in the regions of the continuous magma ocean for different $T_{RF,0}$. Considering the interior composition adds a measurable level of uncertainty since different planets with different atmospheric water content and solidification temperatures can be characterized by the same outgoing OLR. 
Note that unless we are able to constrain the surface pressure of water vapor on an exoplanet, not feasible with the current observational capabilities \citep{Madhusudan2014}, we will not be able to constrain the type of evolutionary (continuous, transient) magma ocean. However, hypotheses and proxies concerning planetary water abundance could break this OLR degeneracy that disappears at low vapor pressures close to 4 bar (see Fig. \ref{fig:GoldblattMulti}). A water-poor planet with a thin atmosphere of 4 bar water would be sensitive to the $T_{RF,0}$ value for developing a continuous/transient MO close to their separation limit. Such could be the case for distinguishing the compositions of HD 219134 b and c if one is found in magma ocean state and the other is not (Fig. \ref{fig:exoplanet}). Kepler 36b' s orbit is further than this distinction possibility and receives enough energy from the star to be in continuous magma ocean as long as it has at least 4 bar water. 
As soon as its atmosphere is lost it would resemble a black body on which the liquid viscosity effect of any water present would ensure the rapid MO solidification, as we showed earlier.
Planets Kepler 236c, Ross 128b and LHS 1140c on the contrary are located in the $F_{lim}$ region for relatively high vapor pressure. Assuming $\geq$200 bar the system converges to the minimum OLR solution of 282 W/m$^2$ (see Fig. \ref{fig:GoldblattMulti}) which is maintained for up to 1000 bar \citep{Ikoma2018}. Detecting any magma ocean state on those planets would be difficult because of the opaque atmosphere. However, if detected it would mean that the planet formed within a water-rich environment that ensured the minimum atmospheric 200 bar required for the continuous magma ocean. Especially for LHS 1140c, the planet LHS 1140b located in the transient MO region of the same system could provide complementary information for the likelihood of high water content. GJ 1132b is located at the compositional distinction limit. Its potential MO has been studied before by \citet{Schaefer2016}. A low atmospheric water content in its MO state would be a proxy of primitive silicate composition. Any of the continuous magma oceans on those planets would eventually solidify if their atmospheric water were lost and were not replenished by the interior. 

The possibility of observing a transient magma ocean system is insignificant due to the order of million years duration that we find for them, which is very short compared to observable systems' ages. Detection of continuous magma oceans on candidate planets (at orbits receiving 282 W/m$^2$ or more (see Fig. \ref{fig:exoplanet}) is challenging but is aided by the fact that the planet's MO brightness temperature would be much higher than that corresponding to its equilibrium temperature, yielding OLR of up to 16,000 W/m$^2$ (see Table \ref{table:Flimits}).  
Such measurements require secondary transit observations as carried out for 55 Cnc e with the Spitzer telescope \citep{Demory2016} aided by the longer wavelength coverage of JWST. A low brightness temperature, in agreement with a low OLR of 282 W/m$^2$, would be an indication towards high steam pressures (See companion paper for possible emission spectra). The surface pressure is not retrievable with the current capabilities but promising methods are developped for low pressure atmospheres (10 bar) that demonstrate pressure broadening of absorbers such as CO$_2$ and O$_2$ \citep{Misra2014}. Transmission methods could not probe high surface pressure atmospheres but the latter's OLR would be already near the runaway greenhouse limit in those cases so one should focus in retrieving the latter. A measured OLR=282 W/m$^2$ would be indicative of MOs with high steam pressures. We suggest the auxiliary/complementary use of observations obtained from the permanent magma ocean type, such as potentially on 55 Cancri e \citep{Demory2016,Angelo2017} and Kepler 78b. From there one could isolate characteristic atmospheric signatures such as: the atmospheric effects of evaporated silicate species that develop over the molten rocky surface \citep{Fegley2016RockSoluble, Kite2016,Hammond2017} and the oxides in the presence of a steam atmosphere \citep{Fegley2016RockSoluble}. Detecting similar silicate cloud signatures on planets close to the continuous MO compositional distinction that is observed at low vapor pressures (4 bar) would serve as a proxy of their composition ($T_{RF,0}$) and of their water content. 

Detection of evolutionary magma oceans additionally requires stellar ages in order to focus on systems with ongoing planetary formation, preferably after recently completed accretion. 
Constraining the albedo from observations is a possibility given favorable orbital configurations \citep{Madhusudan2014,Kite2016} and would help define the range of orbital distances for a conditionally continuous magma ocean.

\section{Discussion}\label{section:discussion}

We previously showed how the MO duration is tied to the outgassing. The latter is sensitive to factors that modify the amount of enclosed melt or the upper mantle temperature. Two such factors are the assumptions of surface rheology front temperature and critical melt fraction. They vary significantly among studies and are sources of deviations when comparing with our Ref-A results \citep[e.g.][$T_{RF,0}$=1370 K]{Hamano2013} \citep[][$T_{RF,0}$=1560 K]{Lupu2014} and \citep[][$\phi_{C}=0.30$]{Hier2017}. 

However, keeping both above assumptions constant, the outgassing in this study still represents an upper limit with respect to other studies. The reason is two-fold.
	Firstly, the use of the one-phase adiabat (Section \ref{section:adiabat}) minimises the amount of enclosed melt at the end of the MO due to its high slope with respect to the melting curves. From the mass conservation follows that the volatile outgassing into the atmosphere maximises. Employing a two phase adiabat instead tends to parallelize the slope to the melting curves and results in more enclosed melt and lower outgassing (e.g. the Ref-A case in \citet{Lebrun2013} outgasses 200 bar H$_2$O compared with 220 bar (this study) via this effect). However, the use of the \citet{Solomatov1993a} two-phase adiabat is subject to strict assumptions (i.e. linear melting curves).

Secondly, we did not account for the depression of the solidus that accompanies the mantle enrichment in water \citep{Katz2003}. Initially, note that the parameterization suggested by \citet{Katz2003} modifies the surface melting temperature $T_{RF,0}$ above the error margin (20 K) for an atmospheric pressure $\ge30$ bar. 
Furthermore, it is only valid for pressures up to 8 GPa, corresponding to a depth of 220-250 km. Indeed, it was motivated by solid state mantle dynamics and explicitly designed to aid modeling of melt generated locally at shallow depth \citep{Noack2012,Tosi2017}. It cannot be extrapolated to higher pressures in the upper mantle, let alone throughout the range of a global MO (covering pressures from the surface down to $~$135 GPa). 

Nonetheless, based on our mass balance (Eq. \eqref{eq:massBalance}), we make a first order estimation of the melting temperature reduction effect during increased water concentration in the melt. Assuming that both the solidus and the liquidus are reduced by the same amount for the same water content (see \citet{Katz2003}, Section 2.2), the MO solidification will take place at a lower temperature. In this respect our model provides lower bounds on the solidification time for the same outgassed quantities (Fig. \ref{fig:evolOutgassingB}). However, estimating the melt fraction using a wet solidus comprises more than a linear shift of melting curves, which would leave the MO final melt fraction unchanged. In fact, the inversion of the saturated solidus appearing near the surface is not necessarily matched in the non-linear shape of the saturated liquidus \citep{Makhluf2017}. 
A wet solidus essentially would increase the enclosed melt at the MO end. Based on our current anhydrous parameterization our final outgassing estimations are upper limits because the remnant melt is here minimum (Fig. \ref{fig:doubleoutgass}). A detailed study is required to quantify the overall effect on the solidification time taking into account the surface solidus depression and the decrease in degassing which exert opposing tendencies on the MO duration. 
 
Factors that decrease the $T_{RF,0}$ (see Table \ref{table:sensitivity}), such as atmospheric steam pressure (1000 bar cause a decrease of 100 K \citep{Katz2003}), melt silicate content (decrease by up to 400 K), and redox state would further increase the solidification time. Significant work has been done towards resolving melt redox evolution \citep[e.g.][]{Schaefer2016,Wordsworth2018} and combining it with silicate content evolution in the melt \citep{Gaillard2015redox}, which is a future step for detailed modeling.

\begin{figure*}[htb!]
	\centering
	\includegraphics[width=0.83\textwidth]{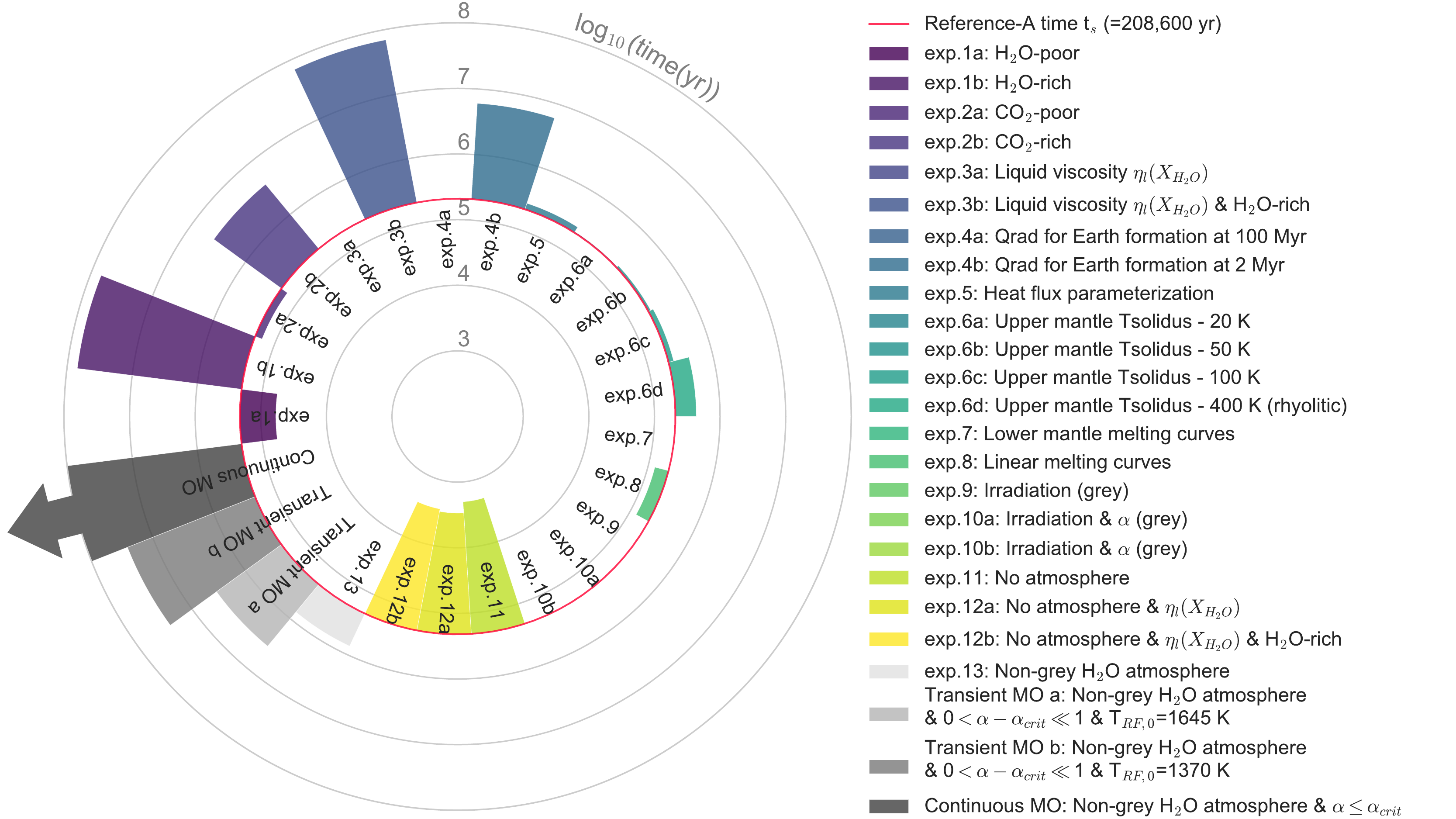}
	\caption{Cumulative plot of the outcome of the sensitivity experiments for the solidification time $t_s$ (in log$_{10}$ units) using the 1D COMRAD model, compared to the Ref-A timescale (red line). Labels are as in Table \ref{table:sensitivity}, which contains the details of each experiment 1--13. Three additional columns are plotted with the outcome of specific settings using the lbl atmosphere (greyscale). Parameters in each experiment as in Ref-A unless otherwise specified. The ``transient'' MO duration corresponds to the $t_s$ obtained for the highest acceptable albedo above $\alpha_{crit}$, to the limit of model resolution (lowest cooling flux 1 W m$^{-2}$), for two different surface rheology front temperatures T$_{RF,0}$. The time arrow of the ``continuous'' MO is obtained for $\alpha\leq\alpha_{crit}$ and hints to effectively unbounded duration, in the absence of atmospheric loss processes.}
	\label{fig:circular}
\end{figure*}


Dynamically, the MO termination is characterized by two main non-linearities. One is the decelerating advance of the solidification front from the bottom upwards that results in a shallow magma ocean of 50 km or less for $\approx 50\%$ of the magma ocean lifetime. The other is the abrupt end of the magma ocean stage, which is marked by a discontinuous viscosity jump of $>$8 orders of magnitude across the critical melt fraction.
The catastrophic H$_2$O outgassing phenomenon is an additional non-linear process. 
For an Earth-sized planet it ensues when the total melt volume fraction drops below 30\% (Fig. \ref{fig:figAspect}). Adopting the Katz parameterization for the late shallow MO stage does not prevent that degree of solidification. This is because even if the solidus depression were to ensure fully molten water--enriched layers, its maximum range of validity is 8 GPa. This barely covers 10\% of the Earth mantle volume. It takes a combination of solidus depression at higher pressures (not yet confirmed) and a two-phase temperature profile such that global melt remains higher than 30\% of mantle volume, in order to hinder the abrupt H$_2$O outgassing (Fig. \ref{fig:figAspect}). Our model shows that initial cooling is instead very rapid and causes the solidification of 90\% of the mantle within few thousands of years via bottom-up crystallization (Fig. \ref{fig:evol3cases}). The phenomenon could be mitigated if solidification proceeded from the middle outwards, maintaining a large part of the mantle molten in the form of a basal MO \citep{Labrosse2007}. A detailed two--phase flow model such as \citet{Hier2017}, expanded to cover the middle point solidification, is required to quantify this effect in detail.

Moreover, we adopt here the two atmospheric species H$_2$O and CO$_2$ but acknowledge the need for including additional trace species that may alter the radiative balance and/or react with surface melt \citep{Gaillard2014,Lupu2014,Zhang2017,Wordsworth2018}. In addition, processes that alter the albedo during the MO evolution (see recent work by \citet{Pluriel2019}) could have an effect on the MO evolution into transient or continuous type. 

The immediate outgassing of CO$_2$ could have an effect on the hydrodynamic escape process which usually is studied on the assumption that CO$_2$ is a minor gas in the atmosphere \citep{Hamano2013,Lupu2014, Hamano2015,Airapetian2017,Wordsworth2018}.
In particular, a low mixing ratio of water in the atmosphere together with abundant CO$_2$ is known to create a cold trap over the convective atmospheric region and to hinder the thermal escape of the heavier and ionized oxygen atoms. 
 \citet{Wordsworth2013CO2richLoss} argue that high CO$_2$ mixing ratio would not effectively prevent the escape. 
 A posterior water regassing process suggested by \citet{Kurokawa2018} could operate through early plate tectonics and could mitigate water loss. It could maintain 2--3 earth oceans bound in the interior against hydrodynamic escape and justify D/H ratios. Combining the response of varying atmospheric composition, a baseline evolution of which we provide here (Fig. \ref{fig:evolOutgassingA}), to different scenarios of early XUV stellar radiation \citep{Lammer2008,Johnstone2015,Airapetian2017,Odert2017} as well as constraining the onset of solid mantle convection \citep{Maurice2017} could help resolve this issue.
 
Lastly, the grey atmosphere is an easily applicable solution for MO modelers but it can be insensitive to the insolation radiation. We also find that it is sensitive to the different absorption coefficient values used for the CO$_2$ (See \citet{Elkins-Tanton2008} for a wide range of $k_{0,CO_2}$ explored for fixed H$_2$O/CO$_2$ atmospheric mixtures). The use of $k_{0,CO_2}$ derived from present ECS studies is unsuitable for the early Earth climate. The lbl approach remains computationally costly, but the pre-calculated OLR values provided in the companion paper for pure steam are a first step towards a wider use of an atmosphere better resolving the absorption in the IR.

\section{Conclusions}\label{section:conclusions}

We have conducted a systematic analysis of the numerous factors and physical processes that affect the thermal and outgassing evolution of a global terrestrial magma ocean (Fig. \ref{fig:circular}). The dominant effect is the steam atmosphere blanketing. Silicate--evolved melts have lower melting temperature which causes linear increase of the solidification time. Such chemical evolution is found to decrease the solidus and it is the next most prominent factor for prolonging the transient MO lifetime. Water dependent viscosity can be ignored for primitive compositions and for planets with greenhouse atmospheres, while it should be considered for atmosphere-free planets and for silicate-evolved melt compositions.

We emphasize that at the end of the magma ocean, the mantle can store between 45 and 10\% of its initial H$_2$O reservoir and only 6\% of the CO$_2$. The massive outgassing of CO$_2$ that precedes the catastrophic H$_2$O outgassing could have an effect in the early atmospheric escape. 
The duration of the magma ocean is closely tied to the degassed amount of volatiles with greenhouse potential. For Earth, its lifetime does not exceed 5 Myr assuming a water reservoir as large as 5 Earth's oceans while CO$_2$ plays a less important role. 

The calculation of the thermal emission for a pure steam atmosphere \citep{Katyal2017} shows that the solidification of the magma ocean can be effectively halted at a suitable minimum surface pressure for a given melting temperature at limits that differ from the constant RG-limit 282 W/m$^2$. Under no combination of parameters is the early Earth found to exist in a continuous magma ocean. 

We find that a molten rocky planet with atmosphere poor in water is a suitable target to acquire information on its mantle surface rheology front temperature. 
The $\sim10,000$ W/m$^2$ difference in OLR for non-massive ($\sim$4 bar) steam atmospheres between planets with and without a magma ocean can be used as a proxy of different melting temperatures that disentangles surface compositions. Surface information would however be masked at higher vapor pressures ($>100$ bar).

We discuss the set of permanent/conditionally continuous/transient MO types. Those can be viewed as stages, among which a planet can be reassigned during stellar evolution or via potential orbital migrations. Future studies on the thermal and chemical evolution of magma oceans in the solar and extrasolar systems can benefit from our comprehensive model analysis of the numerous factors that influence it. In return, our model will benefit from future observations of albedo on exoplanets close to the compositional distinction at low $P_{H_2O}$ OLR limit and spectral properties of permanent magma ocean planets expected from future missions such as \textit{ARIEL} \citep{Turrini2018} and \textit{PLATO} \citep{Rauer2014} [stellar age constraints].

\section*{Acknowledgements}
We thank an anonymous reviewer whose comments helped improve a previous version of the paper, and Melissa McGrath for editorial handling. AN and NT acknowledge financial support from the Helmholtz association (Project VH-NG-1017). NK acknowledges funding from the German Transregio Collaborative Research Centre ``Late Accretion onto Terrestrial Planets (LATP)'' (TRR170, sub-project C5). MG acknowledges financial support from the DFG (Project  GO 2610/1-1). Additional support in the form of conference travel grants for AN was provided from the TRR170 project and the Earth and Life Science Institute in Tokyo. AN wishes to thank Slava Solomatov and Keiko Hamano for insightful conversations.



\appendix

\renewcommand\thefigure{\thesection.\arabic{figure}}
\setcounter{figure}{0} 
\section{Melting curves}\label{appendix:meltCurves}
We report here the fittings that adopted to parameterize the various melting curves used in this study.
\subsection{``Synthetic'' melting curves}\label{appendix:meltingCurveSynthetic}

For the solidus temperature ($T_{sol}$) of nominally anhydrous peridotite and pressures $0 \leq P \leq 2.7$ GPa, we use \citep{Hirschmann2000} :
\begin{equation}
  T_{sol} = 1120.661 + 273.15 + 132.899 P - 5.904 P^2,
  \label{eq:TsHir00}
\end{equation}
with reported error by the authors of $\pm 20$ K. For $2.7 < P \leq 22.5$ \citep{Herzberg2000}:   
\begin{equation}
  T_{sol} = 1086 + 273.15 - 5.7 P + 390 \log(P) \label{eq:TsHerz00},
\end{equation}
with reported error by the authors of $\pm 68$ K. At lower mantle pressures, for $P > 22.5$ GPa, we use a quadratic fit to the data of \citet{Fiquet2010} for fertile peridotite:
\begin{equation}
  T_{sol} = 1762.722 + 31.595P - 0.102P^2. \label{eq:TsFiq10}
\end{equation}

For the liquidus of fertile peridotite, we use a fit to data of \citet{Zhang1994} for $0 \leq P \le 22.5$ GPa:
\begin{equation}
T_{liq} = 2014.497 + 37.743P -0.472 P^2,
  \label{eq:TlZhang00}
\end{equation}
and for $22.5 > P$ GPa, again a quadratic fitting to data of \citet{Fiquet2010}:
\begin{equation}
 T_{liq} = 1803.547 +50.810P -0.185P^2.
  \label{eq:TlFiq10}
\end{equation}

 \subsection{Linear melting curves}\label{appendix:meltingCurveLinear}
 
A linear approximation for the melting curves presented in \citet{Abe97} is used for both solidus and liquidus:
\begin{equation}
 T_{sol} = 1500.0 + \frac{4450.0}{3000.0}z, \quad  T_{liq} = 2000.0 +\frac{4560.0}{3000.0}z
  \label{eq:TsTakahashi93}
\end{equation}
where $z$ is the depth from the surface in km.

\subsection{Andrault melting curves}\label{appendix:meltingCurveAndrault}

The following quadratic fitting to the data of \citet{Andrault2011} for a chondritic composition is employed for the lower mantle only for $P > 22.5$ GPa: 
\begin{equation}
 T_{sol} = 2056.489 + 15.801 P - 0.003P^2, \quad  T_{liq}=2049.555 + 24.671P - 0.035P^2
  \label{eq:TsAndr11}
\end{equation}

Melting curves for the upper mantle are identical to those described above as ``synthetic'', unless otherwise specified.

\section{GREY ATMOSPHERE}\label{appendix:grey}
Assuming optical thickness 1 for a dense atmosphere and evaluating radiative balance at normal optical depth 2/3, \citet{AbeMatsui85} find that the opacity $\tau_i$ for a given species $i$ is proportional to the absorption $k^\prime$: 
\begin{equation}
 \tau_{i}=\frac{3k' P_{i}}{2g},  \label{eq:tauFinal}
\end{equation}
where $P_{i}$ is the partial pressure of the species $i$ in the atmosphere and $k^\prime$ is the absorption coefficient under a certain pressure $P_{i}$. $k^\prime$ is proportional to the atmospheric absorption coefficient $k_{0,i}$ under normal atmospheric conditions ($P_0$, $T_0$) and can be defined as follows:
%
\begin{equation}
 k^\prime=\left(\frac{k_{0,i} g}{3 P_{0}}\right)^{1/2}. \label{eq:kappatau}
\end{equation}

Upon including Eq. \eqref{eq:kappatau} into the opacity relation \eqref{eq:tauFinal}, the opacity $\tau_{i}$ for each volatile is obtained for an atmosphere of pressure higher than the normal conditions $P_0$ (as also  \citet{Pujol2003} and \citet{Elkins-Tanton2008}): 
%
\begin{equation}
 \tau_{i}=\frac{3k M_{i,atm}}{8\pi R_p^2}, \label{eq:tauPre}
\end{equation}
where, $k$ is the absorption coefficient of the volatile at the surface, $R_p$ the planetary radius, and $M_{i,atm}$ the mass of the volatile $i$ in the atmosphere.

In the grey approximation, the total opacity of the atmosphere ($\tau$) is given by the sum of the opacities of each gas, i.e. $\tau=\Sigma_i\tau_i$ \citep{Pujol2003,Elkins-Tanton2008}. The opacity is a measure of the radiative absorption through atmospheric layers and is inversely proportional to their emissivity $\epsilon$. Following \citet{AbeMatsui85}, the two quantities are linked as follows:
\begin{equation}
 \epsilon=\frac{2}{2+\tau}.
 \label{eq:tauemm}
\end{equation}
%

The atmosphere is assumed to be in radiative-convective equilibrium and the TOA is defined to occur at the base of the stratosphere, above which the temperature is governed by pure radiative balance. The assumptions include the plane-parallel approximation for the atmospheric layers and ignore radiative contributions from directions wider than 60$^\circ$ degrees between neighboring layers. More information on the derivation of the above equations can be found in \citet{AbeMatsui85}.

\section{LINE-BY-LINE MODEL DATA}\label{appendix:LBL}

\subsection{Lbl atmospheric data product}
\begin{figure*}[htb!]
\centering
\includegraphics[width=0.65\textwidth]{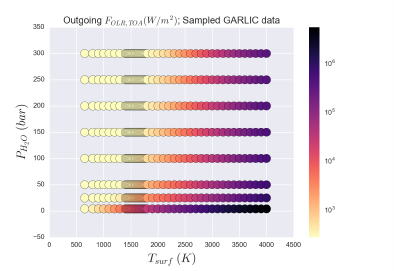}
\caption{OLR$_{TOA}$ sampled on the $(T_{surf}$, $P_{H_2O})$-space obtained using the lbl model. The detailed calculation of the values is found in the companion paper.}
\label{fig:Product}
\end{figure*}

In Fig. \ref{fig:Product} we show the OLR on the $50\times 8$ grid of $\left(T_{surf}, P_{H_2O}\right)$ points, that we used as input for our  simulations. The OLR data at each grid point have been obtained with the method described in the companion paper using a line-by-line code (GARLIC) of \citet{Schreier2014}. 

The grid spans surface temperatures from 650 to 4000 K and water vapor surface pressures from 4 to 300 bar. It is irregularly spaced and denser over the temperature range where the highest rate of enrichment and outgassing takes place. This range is obtained by performing simulations using synthetic melting curves with the interior model coupled to the grey atmosphere model (see Fig. \ref{fig:figAspect}). For $T_s \in \left[1400, 1800\right]$ K, the OLR was sampled with a resolution of 20 K, while a 100 K resolution was employed outside this range. 
The sampling is sparser (8 values) on the pressure axis for $P\in \left[4,300\right]$ bar. 
In order to obtain the OLR values at intermediate $(P,T)$ points of the above dataset, a bilinear interpolation method was used \citep{Python2001}. 
In order to estimate the interpolation error, we compared the interpolated field with an independent set of intermediate data points obtained from the atmospheric model. The relative interpolation error amounts between 1 and 10\%. The maximum of 10\% occurs at pressures lower than 10 bar and high temperatures. The minimum occurs for high pressures and temperatures at the lower end of the dataset. Therefore, the quality of the result is acceptable for this study that focuses on the coolest end of magma ocean phase where the outgassed atmosphere has high pressure and the errors are minimal (1-2 W/m$^2$). 


The data of Fig. \ref{fig:Product} represent the OLR at the TOA with a viewing angle of $38^\circ$ and thus differ from the field shown in Fig. \ref{fig:greyVSLBL}, which represents the net planetary flux at the TOA. More details on the OLR value calculation can be found in the companion paper. 

In order to satisfy the requirement of our iteration algorithm for surface temperatures lower than $T_{H2O,crit}=647$ K, which are not covered by our gird, we use a fit to the OLR data of \citep{Naka92}.
This aspect does not affect our results for the solidification process, which occurs for $T_{surf} \approx T_{RF,0} \gg T_{H2O,crit}$, but ensures that the iteration algorithm runs unhindered until convergence to the solution.


\subsection{Limiting radiation values}

We use Eq. \eqref{eq:albedoMax} to estimate the orbital distance for which a planet of given albedo is located at the boundary that separates a long-term and short-term magma ocean. To this end, the value of the limiting radiation $F_{lim}$ corresponding to a specific water vapor pressure $P_{H_2O}$ and rheology front temperature at the surface $T_{RF,0}$ is needed.
In Table \ref{table:Flimits}, we report $F_{lim}(P_{H_2O},T_{surf})$ for two different rheology front surface temperatures as obtained by interpolating the OLR data points of the companion paper (Fig. \ref{fig:Product}). The same values are plotted in Fig. \ref{fig:GoldblattMulti} and used to calculate the critical distances for the young Sun in Fig. \ref{fig:albedoMulti}.

\begin{table}[htb]
\addtolength{\textfloatsep}{0.2mm}

\caption{$F_{lim}(P_{H_2O}, T_{RF,0})$ for indicative $T_{RF,0}$ cases calculated with the \citet{Katyal2017} data.}
\centering
\begin{tabular}{c c c}
\hline
\hline
  $P_{H_2O}$ (bar)  & $F_{lim}$($T_{RF,0}=1645$ K)  (W/m$^2$) & $F_{lim}$($T_{RF,0}$=1370 K) (W/m$^2$)  \\
  \hline
   4 &  16724.2 & 5885.6 \\ 
   25 & 1362.8  & 519.4 \\ 
   50 &  677.3  & 332.8 \\
   100 &  398.5 & 286.6  \\
   200 & 298.1   & 283.6  \\
   300 & 285.8  & 283.6  \\ 
    \hline
 \multicolumn{3}{l}{}
 \end{tabular}
 \label{table:Flimits}
 \end{table}

\section{Radioactive decay}\label{section:Qrad}

\begin{table}[htbp]
\addtolength{\belowcaptionskip}{0.5mm}
\caption{Parameters used in Eq. \eqref{eq:qrad} to compute the radiogenic heat production.}
\centering
\begin{tabular}{c c c c}
\hline
\hline
Radioactive isotope       & Concentration $X^0$  & Heat production $Q$ (W/kg) & Half-life $\lambda$ (yr) \\
\hline
$^{238}$U (*)          & $6.23\cdot 10^{-8}$  & $9.46\cdot 10^{-5}$  & $4.47\cdot 10^{9}$ \\
$^{235}$U (*)          & $1.97\cdot 10^{-8}$  & $5.69\cdot 10^{-4}$  & $0.704\cdot 10^{9}$  \\  
$^{40}$K  (*)          & $4.61\cdot 10^{-7}$  & $2.94\cdot 10^{-5}$  & $1.25 \cdot 10^{9}$ \\
$^{232}$Th (*)         & $1.54\cdot 10^{-7}$  & $2.54\cdot 10^{-5}$  & $14.5 \cdot 10^{9}$  \\ 
$^{26}$Al ($\dagger$)  &$1.23\cdot 10^{-6}$  & 0.455      & $0.717 \cdot 10^{6}$ \\
$^{60}$Fe ($\dagger$)  &$7.2\cdot 10^{-10}$  & 0.0412               & $2.62 \cdot 10^{6}$  \\
\hline
\multicolumn{4}{l}{(*) Parameter values are from \citet{Schubert2001}.}\\
\multicolumn{4}{l}{($\dagger$) Concentrations from \citet{McDonough1995}. Heat productions and half lives from \citet{Neumann2014}.}
\end{tabular}
\label{table:Qrad} 
\end{table}

The heat production due to the long-lived radioactive elements $^{238}$U, $^{235}$U, $^{40}$K and $^{232}$Th, and the short-lived elements $^{26}$Al and $^{60}$Fe is taken into account in the energy balance equation \eqref{eq:energy} via the term $q_r$ whose explicit expression reads:
\begin{equation}
 q_r = \sum_i X^0_i Q_i \exp\left(-\ln(2)\frac{t+t_0}{\lambda_i}\right),
\label{eq:qrad}
\end{equation}
where, for each element $i$, $X^0_i$ the isotope concentration in the silicate mantle at the formation time of the CAI (4.55 Gyr ago), $Q_i$ the specific heat production, $\lambda_i$ the half-life, $t_0$ the assumed formation time of the magma ocean (e.g. 2 or 100 Myr after the CAI as in experiment 4 in Table \ref{table:sensitivity}), and $t$ the time (with $t>t_0$). For the long-lived elments, the initial isotope concentration $X^0_i$ is calculated by scaling back in time its present-day  concentration according to the isotope half-life. In Table \ref{table:Qrad}, we report for each isotope the parameters of Eq. \eqref{eq:qrad}. The energy released by the decay of the radioactive isotopes is made available to the whole magma ocean volume.

\section{EXOPLANETS}\label{appendix:exoplanets}


\begin{table}[ht]
\addtolength{\belowcaptionskip}{1mm}
\caption{Planet and host star parameters used in Fig. \ref{fig:exoplanet}}
\hspace{-2.8cm}
\begin{tabular}{l c c l c c c cc}
\hline
\hline
Planet  & Orbital distance R & Host star &  T$_{eff}$& R$_{star}$ & Luminosity L & Ref. \\
& (AU) & & (K) & (in solar units R$_{\astrosun}$) & (in solar units L$_{\astrosun}$) & \\  
\hline
$^{*}$Kepler 236 c & 0.1320 &  Kepler 236 & 3750 & 0.510    &  -  &  exoplanet.eu/catalog/ \\ 
55 Cancri e &  0.0156   & 55 Cancri & - & - & 0.59 &  exoplanet.eu/catalog/ \\
Earth &    1.0000      & Sun (100 Myr old) &  5326  & 1.000 &0.72 & \citet{Gough1981} \\
$^{*}$Kepler 36 b &  0.1151   &  Kepler 36 & 5911 & 1.619 & - & openexoplanetcatalogue.com \\
$^{*}$HD 219134 b\&c & 0.0388 \& 0.0653   & HD 219134 & 3131 &0.186 & - &  exoplanet.eu/catalog/ \\
$^{*}$Ross 128 b &  0.0496   &  Ross 128 & 3192 & 0.197 & - & openexoplanetcatalogue.com \\
$^{*}$GJ 1132 b & 0.0154   &  GJ 1132 & 3270 & 0.207 & - & openexoplanetcatalogue.com \\ 
$^{*}$LHS 1140 b\&c & 0.087 \& 0.02675   & LHS 1140 & 4699 & 0.778 & - &  openexoplanetcatalogue.com \\  
$^{*}$Kepler 78 b &  0.01  & Kepler 78  & 5089 & 0.74 & - & exoplanet.eu/catalog/ \\                   

\hline
\multicolumn{7}{l}{$^{*}$In absence of values of luminosity relative to $L_{\astrosun}$ we calculated the stellar luminosity directly from $T_{eff}, R_{Star}$ data.}
\end{tabular}
\label{table:exoplanets}
\end{table}
 
\section{CALIBRATION OF THE MELT VISCOSITY PREFACTOR}\label{appendix:viscCalibration}

The composition of anhydrous peridotite, which we employed to define our ``synthetic'' melting curves (a list of the corresponding oxides can be found in \citet{Hirschmann2000}), is not covered by the empirical model of \citet{Giordano2008} that we used to determine the liquid viscosity and its dependence on the water concentration (\ref{section:viscosity}). 
However, we found that the composition of basanite that belonged to the \citet{Giordano2003,Giordano2008} model calibration database is able to reproduce the temperature-dependent viscosity values of anhydrous silicate obtained experimentally \citep{Urbain82}, within less than 10\% relative error. To be consistent with the assumption of a peridotitic composition, it is important to use a composition as close to a primitive one as possible. Indeed the composition of basanite is among the least evolved in the classification of melts \citep{LeBas1986}.

Assuming such composition and fitting the model of \citet{Giordano2008} to the experimental data of \citet{Urbain82} (see Fig. \ref{fig:viscosityVFT}), we obtained a modified prefactor in Eq. \eqref{eq:nliqGiordano}, namely $A_G = -3.976$. The result is within the acceptable range of $A_G =-4.55\pm 1$ log unit, given by the model authors \citep{Giordano2008}.
As shown in Table \ref{table:nliq_calibration}, the use of this prefactor yields an error relative to the experimental values smaller than 10\%. Note that we use this calculation only to provide a first-order estimate of the effects of water on the melt viscosity without needing to explicitly describe the evolution of the melt composition, which is beyond the scope of the present work.

\begin{table}[htbp]
\addtolength{\belowcaptionskip}{0.2mm}
\caption{Comparison between values of the viscosity of anhydrous liquid peridotite obtained experimentally and calculated with the model of \citet{Giordano2008} assuming a basanite composition and a prefactor $A_G=-3.976$ in Eq. \eqref{eq:nliqGiordano}.}
\centering
\begin{tabular}{l c c c}
\hline
\hline
$T$ (K) & Experimental $\eta_l$ (Pa s) &  Calculated $\eta_l$ (Pa s) & Error (\%)    \\
\hline
2000  & 0.22 &   0.2350  &  6.84  \\ 
2220 &  0.08 &   0.0788  &  $-1.47$  \\
2300 &  0.06 &   0.0579  &  $-5.01$ \\
\hline
\end{tabular}
\label{table:nliq_calibration}
\end{table}


\section{Constants}\label{appendix:Constants}

Table \ref{tableConstants} includes the constants and parameters used in most of the simulations unless otherwise stated.


\begin{table}[htb]
\addtolength{\belowcaptionskip}{0.2mm}
\addtolength{\abovecaptionskip}{0.2mm}
\caption{Model constants and parameters.}
\centering
\begin{tabular}{l c c c}
\hline
\hline
  Parameter  & Value & Unit  & Description  \\
 \hline
   $R_p$           &  6371     & km        & Planetary radius\\ 
   $R_b$           &  3481     & km        & Core mantle boundary radius\\  
   $g$             & 9.81      & ms$^{-2}$ & Gravity acceleration\\
   $k_{0,H_2O}$  &  0.01     & m$^2$/kg        & Absorption coeff. at $P_0$, $T_0$\\ 
   $k_{0,CO_2}$  &  0.001    & m$^2$/kg        & Absorption coeff. at $P_0$, $T_0$ \\ 
   $P_0$           & 101325    & Pa        & Normal atmospheric pressure for $k_{0,vol}$ \\
   $T_0$           & 300       & K        & Normal atmospheric temperature for $k_{0,vol}$ \\
   $\alpha_0$      & 3 $\cdot$ 10$^{-5}$   & K$^{-1}$ & Mantle thermal expansivity \\
   $K_0$           & 200       & GPa       & Mantle bulk modulus \\
   $K^\prime$      &  4        & --        & $P$-derivative of mantle bulk modulus \\
   $m$             &  0        &  --       & Parameter in Eq. \eqref{eq:aTofP} \\
   $c_P$           & 1000      & J kg$^{-1}$K$^{-1}$ & Mantle isobaric thermal capacity \\
   $\kappa_T$      & 10$^{-6}$ & m$^2$s$^{-1}$  & Mantle thermal diffusivity \\
   $k_T$           & $\kappa_T/(\rho c_P)$  & J K$^{-1}$s$^{-2}$m$^{-2}$ & Mantle thermal conductivity \\
   $\rho_l$        & 4000      & kg m$^{-3}$ & Melt density\\
   $\rho_s$        & 4500      & kg m$^{-3}$ & Solid density\\
   $\phi_C$        & 0.4       &  --         & Critical melt fraction \\
   $\eta_0$        & $4.2\cdot 10^{10}$ & Pa s & Solid viscosity prefactor \\
   $E$             & 240        & kJ mol$^{-1}$            & Activation energy \\
  $V$             & 5          & cm$^3$ mol$^{-1}$        & Activation volume\\
   $R$             &  8.314     & J mol$^{-1}$ K$^{-1}$ & Ideal gas constant \\    
   $A_{G}$         &  3.9759    & -- & Prefactor in Eq. \eqref{eq:nliqGiordano} calibrated for basanite \\
   $B_{G}$         &  $^{*}$ & K  & Parameter in Eq. \eqref{eq:nliqGiordano} \\
   $C_{G}$         &  $^{*}$ & K  & Parameter in Eq. \eqref{eq:nliqGiordano}\\
   $A_{K}$         & 0.00024    & -- & Prefactor for hydrous liquid in Eq. \eqref{eq:nliqKarki}  \\
   $B_{K}$         &  4600      & -- & Parameter for hydrous liquid in Eq. \eqref{eq:nliqKarki} \\
   $C_{K}$         &  1000      & K  & Parameter for hydrous liquid in  Eq. \eqref{eq:nliqKarki} \\
   $\kappa_{H_2O,pv}$   & $1.0 \cdot 10^{-4}$ & -- & H$_2$O partition coeff. in solid perovskite\\  
   $\kappa_{H_2O,lhz}$ & $1.1 \cdot 10^{-2}$ & -- & H$_2$O partition coeff. in solid lherzolite \\
   $\kappa_{CO_2,pv}$   & $5.0 \cdot 10^{-4}$ & -- & CO$_2$ partition coeff. in solid perovskite \\
   $\kappa_{CO_2,lhz}$  & $2.1 \cdot 10^{-3}$ & -- & CO$_2$ partition coeff. in solid lherzolite \\
 \hline
 \multicolumn{4}{l}{$^{*}$The value of this parameter is dynamically calculated during the simulation.}
 \end{tabular}
 \label{tableConstants}
 \end{table}


\begingroup
\let\clearpage\relax
\bibliographystyle{aasjournal}
\bibliography{bibAugustPaper-new.bib}
\endgroup

\end{document}